\documentclass[graybox]{SNmult}

\usepackage{type1cm}        

\usepackage{makeidx}         
\usepackage{graphicx}                                   
\usepackage{multicol}     
\usepackage[bottom]{footmisc}
\usepackage{macros}

\usepackage{newtxtext}     
\usepackage[varvw]{newtxmath}      
\usepackage{tikz}
\usepackage{pgfplots}
\usepackage[export]{adjustbox}
\usetikzlibrary{spy}

\usepackage{hyperref}
\usepackage{xcolor}
\definecolor{darkgreen}{rgb}{0.0, 0.5, 0.0}
\definecolor{darkblue}{rgb}{0.0, 0, 1.}

\makeindex

\begin{document}

\title*{Learning  spatially varying regularisation parameters of low regularity for image reconstruction
}
 \titlerunning{Learning   regularisation parameters of low regularity for imaging}
\author{Kostas Papafitsoros, Luca Calatroni, 
Andreas Kofler}
\authorrunning{}
\institute{Kostas Papafitsoros \at School of Mathematical Sciences, \\Queen Mary University of London,\\ London E1 4BS, United Kingdom\\ \email{k.papafitsoros@qmul.ac.uk}
\and Luca Calatroni \at MaLGa centre, DIBRIS, Università di Genova\\
Istituto Italiano di Tecnologia, Genoa, Italy\\\email{luca.calatroni@unige.it}
\and Andreas Kofler \at Physikalisch-Technische Bundesanstalt (PTB),\\ Braunschweig and Berlin, Germany \\\email{ andreas.kofler@ptb.de}
}

\maketitle

\abstract{
In this chapter, we review and discuss the regularity properties of spatially adaptive regularisation weight functions used in variational image reconstruction. Incorporating such weights into classical model-based regularisers, such as Total Variation (TV) and Total Generalised Variation (TGV), allows the regularisation strength to vary across the image and adapt to local image content. When appropriately estimated, these weights can thus significantly improve edge and detail preservation in the reconstructions. We review the existing theoretical literature on this topic for different regularity classes, including constant, continuous, and piecewise constant functions. Our discussion is motivated by recent work on hybrid image reconstruction methods that combine model-based regularisation with deep neural networks to learn highly adaptive regularisation weights. In particular, we discuss how the structural properties of these weights influence the reconstruction from both theoretical and practical perspectives. Through representative examples in image denoising and magnetic resonance imaging (MRI) reconstruction, we demonstrate that the learned weights are often of low regularity and can adapt not only to the image structure but also to the specific noise realisation. We conclude by highlighting several directions for future research on this topic.
}

\section{Introduction}
\label{sec:introduction}
\begin{figure}[t]
\centering
\includegraphics[width=0.97\textwidth, trim={0.5cm 0 6.5cm 0},clip]{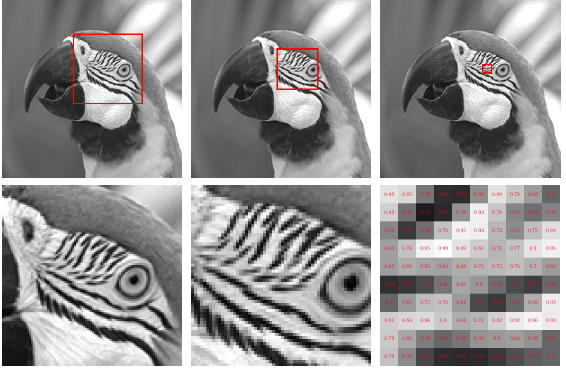}
\caption{Intensity values for a 
grayscale digital image over a pixel grid, where a value, typically in $[0,1]$, represents the brightness of the pixel (0 for black, 1 for white). We regard these images as discretised versions of functions from a domain $\om\subset \RR^{2}\to \RR$.}
\label{fig:pixels}       
\end{figure}

This chapter deals with inverse imaging problems that denote the task of recovering an image $u_{\mathrm{true}}$ from a corrupt and noisy version. Mathematically, we assume that we are given some $f\in Y$ that satisfy the equation
\begin{equation}\label{intro:general_eq}
f=Au_{\mathrm{true}}+\eta.
\end{equation}
Here $u_{\mathrm{true}}\in X$ is the image we would like to recover,  $A: X\to Y$ is a bounded linear forward operator, and $\eta\in Y$ is a random noise component. Furthermore, $X$, $Y$ denote appropriate Banach spaces, with  $X$ typically being a function space consisting of functions $u:\om \subset \RR^{d} \mapsto \RR$. The latter space models digital images, with pixels corresponding to points $x\in \om$, see Fig.~\ref{fig:pixels}. In this paper, we consider only grayscale images, so that the intensity $u(x)$  of the image at each point $x\in\Omega$ is $\RR$-valued. We note that,  even though typically $d=2$ or $3$, corresponding to  2D or 3D images respectively, many theoretical results can be posed in arbitrary dimension. 

Depending on the specific application at hand, the forward operator $A$ can be very simple, e.g.\ $A=Id$ for denoising, or more involved; in Magnetic Image Reconstruction, $A$ performs subsampling of the Fourier coefficients of the signal encoding the imaged tissue. In general, $A$ can be non-invertible or have an unbounded inverse. Because of this, the unknown noise component gets amplified upon (generalised) inversion. More precisely, the task of retrieving $u_{\mathrm{true}}$ given $f$ is an \textit{ill-posed inverse problem}. A standard approach to compute an approximation of $u_{\mathrm{true}}$ is to resort to a variational regularisation approach \cite{Benning_Burger_2018, scherzer2009variational}, i.e.\, solving a problem of the following type:
\begin{equation}\label{intro:general_min}
\min_{u\in X} \; \mathcal{D}(Au,f)+ \mathcal{R}(u;\mathrm{\LL}),
\end{equation}
where $\mathcal{D}$ denotes the \textit{data fidelity term} that ensures that a  minimiser of \eqref{intro:general_min} (the \textit{reconstruction}) will be consistent with the data. On the other hand, $\mathcal{R}$  denotes the \textit{regularisation term} (or \textit{regulariser} which makes the problem well-posed. It typically provides the solutions to \eqref{intro:general_min} with additional regularity with respect to the one characterising the data $f$, and it is largely responsible for the quality and stability of the reconstruction. Finally, $\Lambda>0$ denotes the regularisation parameter (set of scalars or a function) that balances the effect of the regulariser and the fidelity term. Different values of $\Lambda$ impose different regularisation strength, strongly influencing the structure of the solution of \eqref{intro:general_min}. Traditionally, the regularisation parameter is considered to be  a constant function (e.g., a scalar), so that the latter effect is uniform across the image domain. In the following, we will be using the notation $\lambda$  when the regularisation parameter is scalar, or a set of scalars in general, belonging to $\RR_{+}^{\ell}$, with $\ell$ denoting the number of components of the regulariser. In the case of a non-constant parameter map, we will use instead the notation $\Lambda:\om \to (\RR_{+})^{\ell}$. In that case, the regularisation parameter is spatially varying, or even spatio-temporally varying for time-dependent problems. 

\begin{svgraybox}
Spatially varying regularisation parameters are used to weight regularisation with different strength throughout the spatial domain, aiming at sharper structures and better detail preservation. This chapter discusses theoretical aspects regarding the functional \emph{regularity} of $\Lambda$, the structure of the solutions of the corresponding variational problems, and how these can be linked to modern data-driven approaches used to efficiently choose them in practice by means of deep neural networks.
\end{svgraybox}

\section{Model-based regularisation functionals}

In what follows, we assume that  $\eta$ corresponds to a realisation of a Gaussian random variable. In the function space setting, this is typically modelled by a function of low regularity, e.g.\ $\eta\in L^{2}(\om)$. In that case, the choice of a  $L^{2}$-fidelity term arises naturally from the Maximum A Posteriori estimation \cite{kaipio2005statistical}, and  \eqref{intro:general_min} reads 

\begin{equation}\label{general_min_L2}
\min_{u\in X}\; \frac{1}{2}\|Au-f\|_{L^2(\Omega)}^{2} + \mathcal{R}(u;\mathrm{\LL}).
\end{equation}

We next briefly review popular model-based (handcrafted) regularisation functionals based on derivatives. Unless it is otherwise specified, the image domain $\om$ is an open, bounded, Lipschitz domain of $\RR^{d}$.

\begin{figure}[t]
\centering
\begin{minipage}[t]{0.24\textwidth}
\centering
\includegraphics[width=0.97\textwidth]{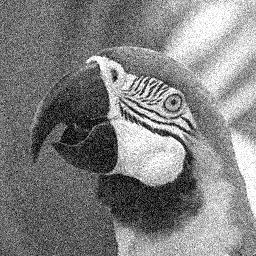}
\end{minipage}
\begin{minipage}[t]{0.24\textwidth}
\centering
\includegraphics[width=0.97\textwidth]{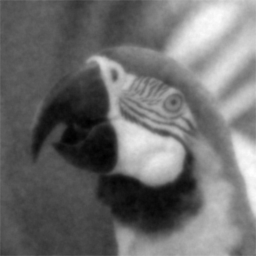}
\end{minipage}
\begin{minipage}[t]{0.24\textwidth}
\centering
\includegraphics[width=0.97\textwidth]{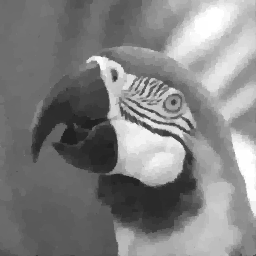}
\end{minipage}
\begin{minipage}[t]{0.24\textwidth}
\centering
\includegraphics[width=0.97\textwidth]{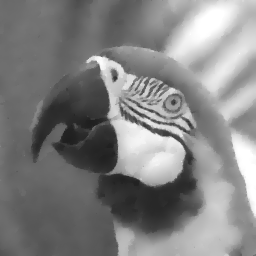}
\end{minipage}

\begin{minipage}[t]{0.24\textwidth}
\centering
Noisy
\end{minipage}
\begin{minipage}[t]{0.24\textwidth}
\centering
Tikhonov
\end{minipage}
\begin{minipage}[t]{0.24\textwidth}
\centering
Total Variation
\end{minipage}
\begin{minipage}[t]{0.24\textwidth}
\centering
Total Generalised Variation
\end{minipage}

\caption{Comparison of classical derivative-based regularisers for a Gaussian denoising task. Tikhonov regularisation produces smooth outputs, destroying the edges; TV preserves edges but also leads to staircasing artifacts, while TGV both preserves edges and eliminates the staircasing artifacts.}
\label{fig:intro_comparison}      
\end{figure}

\subsection{Tikhonov}
Motivated by the idea of reducing oscillations in the image, one of the simplest choices is to choose $ \mathcal{R}(u;\lambda)=\lambda\|\nabla u\|_{L^{2}(\om)}^{2}$, leading to the so-called Tikhonov (or Tikhonov/Sobolev) regularisation \cite{Tikhonov}. In that case, and for a $A:L^{2}(\om)\to L^{2}(\om)$, solutions naturally belong to the Sobolev space $H^{1}(\om)$ \cite{adams2003sobolev} and the problem \eqref{general_min_L2} reads 
\begin{equation}\label{min_tikhonov}
\min_{u\in H^{1}(\om)}\; \frac{1}{2}\|Au-f\|_{L^2(\Omega)}^{2} + \lambda \int_{\om}|\nabla u|^{2}dx.
\end{equation}
This simple approach produces smooth reconstructions and destroys edges, see Fig.~\ref{fig:intro_comparison}. On a theoretical level, this can also be seen from the embedding $\mathcal{C}(\overline{\om})\subset H^{1}(\om)$ that holds for $d=2$: no jump discontinuities can thus occur for a minimiser of \eqref{min_tikhonov}.

\subsection{Total  Variation}
Total Variation (TV) regularisation \cite{ChambolleLions, rudin1992nonlinear} is a classical regulariser based on first-order information but able to preserve image discontinuities, that is, edges. For $\lambda>0$, it is here defined as $\tv_{\lambda}(u): L^{1}(\om) \to \RR\cup\{+\infty\}$ with
\begin{equation}\label{TV}
\tv_{\lambda}(u)=\sup \left \{ \int_{\om} \di \phi\,dx: \phi\in \mathcal{C}_{c}^{\infty}(\om;\RR^{d}), \|\phi\|_{\infty}< \lambda\right \}.
\end{equation}
It can be shown that for  $u\in L^{1}(\om)$, $\tv_{\lambda}(u)<\infty$ if and only $u\in \bv(\om)$, the space of functions of bounded variation \cite{AmbrosioBV} and in that case $\tv_{\lambda}(u)=\lambda|Du|(\om)=\lambda \int_{\om}d|Du|$, where $Du$ is the finite Radon measure that represents the distributional derivative of $u$ and $|Du|$ is its total variation measure. Note that if $u\in W^{1,1}(\om)$, then $\tv_{\lambda}(u)=\lambda \int_{\om} |\nabla u|\, dx$, the $L^1$ norm of the weak gradient of $u$. In general, %
the minimisation \eqref{general_min_L2} reads in this case
\begin{equation}\label{min_TV}
\min_{u\in \bv(\om)}\; \frac{1}{2}\|Au-f\|_{L^2(\Omega)}^{2} + \lambda \int_{\om}d|Du|.
\end{equation}
Problem \eqref{min_TV} has been extensively studied with the idea of understanding the regularising mechanism of TV and the fine-scale structure of its solutions. We note that the majority of the results outlined below concern the denoising case, i.e.\ $A=Id$.\\[0.5em]

\noindent
\textbf{No new discontinuities are created}: In the seminal work \cite{caselles2007discontinuity}, it was shown that if $f\in \bv(\om)\cap L^{\infty}(\Omega)$, then the solution $u$ of the $L^{2}$-TV denoising problem have no other jump discontinuities than the ones already present in the data $f$, up to a $\mathcal{H}^{d-1}$-null set, i.e. $\mathcal{H}^{d-1}(J_{u}\setminus J_{f})=0$, where $\mathcal{H}^{d-1}$ denotes the ($d-1$)-dimensional Hausdorff measure. Here $J_{u}$ and $J_{f}$ denote the jump sets of $u$ and $f$, respectively; see \cite{AmbrosioBV} for precise definitions. 
This result was also proven in \cite{valkonen2015jump} using a different technique and in a slightly more general setting ($L^{p}$ fidelity, $1<p<\infty$, and including the Huber version of TV). In \cite{caselles2007discontinuity}, it was also proven that the magnitude of jump of the solution is smaller than the one of the data almost everyewhere on the jump set of $u$. The jump set inclusion was also proven in \cite{lasica2021existence} for a more general class of regularisers that satisfy a mild regularity assumption. More refined results can be shown in the one-dimensional case, see e.g.\ \cite{cristoferi2026monotonicity, ring2000structural, analyticalaspects_2017}.\\
\textbf{The staircasing effect}: Staircasing denotes the creation of locally constant areas in the minimising solution, a typical feature of TV, see Fig.~\ref{fig:intro_comparison}.  In dimension one, the solution $u$ is constant in areas where $u\ne f$ (in an appropriate almost everywhere sense) \cite{ring2000structural, analyticalaspects_2017} and also close to the boundary of $\om$. Weaker results exist in higher dimensions; for instance, in \cite{jalalzai2016some} it was shown that the solution is locally flat at all global extrema of the data and at all extrema of the solution. We refer the reader to \cite{chambolle2016total} for a more refined characterisation of the areas of the solution that suffer from the stairacasing effect.

\subsection{Total Generalised Variation}
Total Generalised Variation (TGV) was introduced in \cite{TGV} as a higher-order extension of TV, which preserves edges but also has the capability of eliminating the staircasing effect, see Fig.~\ref{fig:intro_comparison}. For $u\in \bv(\om)$, TGV is defined as 
\begin{equation}\label{TGV}
\tgv_{\lambda_{0}, \lambda_{1}}(u)=\min_{w\in \mathrm{BD}(\om)} \lambda_{1} \int_{\om} d|Du-w|+ \lambda_{0} \int_{\om} d|\mathcal{E}w|.
\end{equation}
Here, $\mathcal{E}$ denotes the measure that represents the distributional symmetrised gradient of $w\in \mathrm{BD}(\Omega)$, the space of functions of bounded deformations \cite{bd}. The two regularisation parameters $\lambda:=(\lambda_{0}, \lambda_{1})$ balance not only the effect of the TGV regulariser and the fidelity term but also control the interplay of the two terms in the minimisation in \eqref{TGV}. Observe that if the minimiser $w=0$ in \eqref{TGV} then the functional is equal to TV, whereas if $w=\nabla u$ (the absolutely continuous part of $Du$ with respect to the Lebesgue measure), then TGV behaves roughly like second-order TV (the total variation of the Hessian). We recall in this regard the following proposition from \cite{Papafitsoros_Valkonen_2015}, which states that if the ratio $\lambda_{0}/\lambda_{1}$ is high enough then TGV essentially behaves like $\tv$:

\begin{proposition}
[from \cite{Papafitsoros_Valkonen_2015}]\label{prop:tgv_large_ratio}
There exists a constant $C>0$ depending only on the domain $\om$ such that if $\lambda_{0}, \lambda_{1}>0$ satisfy $\lambda_{0}/\lambda_{1}>C$ then
\begin{equation}\label{tgv_large_ratio}
\tgv_{\lambda_{0}, \lambda_{1}}(u)=\lambda_{1} |Du-m_{\mathcal{E}}(\nabla u)|(\Omega), \quad \text{for all }u\in \bv(\om),
\end{equation}
where for  $v\in L^{1}(\om,\RR^{d})$ we define
$m_{\mathcal{E}}:=\operatorname{argmin}\,\{\|v-w\|_{L^{1}(\om;\RR^{d})}:\, w\in Ker \mathcal{E}\}$.
Recall that $Ker \mathcal{E}=\{r(x)=Bx+c:\;c\in\mathbb{R}^{d},\,B\in \mathbb{R}^{d\times d} \text{ skew symmetric}\}$.
\end{proposition}
The analogous  problem to \eqref{min_TV} now reads
\begin{equation}\label{min_TGV}
\min_{u\in \bv(\om)}\; \frac{1}{2}\|Au-f\|_{L^2(\Omega)}^{2} + \tgv_{\lambda_{0}, \lambda_{1}}(u),
\end{equation}
and as in the TV case, several studies have looked into structural properties of its solutions, mostly for denoising. The one-dimensional case is well understood \cite{bredies2012properties, papafitsoros2015study}: in that case, one can easily show that the solution $u$ is piecewise affine in areas where $u\ne f$. The inclusion of discontinuities also holds in dimension one, which has also been shown in higher dimensions but only under some (mild) assumptions, namely boundedness of the solution and under some smoothing of the TGV functional or mild modifications of it \cite{chambolle2024inclusion, valkonen2017jump}.

In the context of imaging problems, TV and TGV are the most studied and understood handcrafted regularisers and, as such, the ones that have been mainly extended to versions with spatially varying regularisation parameters. In the following, we thus focus our attention on these two models, while, of course, a similar approach can be used for Tikhonov regularisation too \cite{gazzola2025automatic}. We further mention that the literature on model-based regularisers is much vaster, see for instance, higher-order ones \cite{bredies2020higher, papafitsoros2015novel}, employing $L^{p}$ norms \cite{journal_tvlp}, Hessian-based regularisers \cite{lefkimmiatis2013hessian} or anisotropic ones \cite{parisotto2020higher}, just to name a few.

\subsection{Spatially varying regularisation parameters}

We now turn our attention to versions of TV and TGV that employ spatially varying regularisation parameters to favour local adaptivity. Denoting, respectively, by $\Lambda: \om \to \RR_{+}$  and $\Lambda: \om \to (\RR_{+})^{2}$ with $\Lambda=(\Lambda_{0}, \Lambda_{1})$, we thus consider:
\begin{align}
\tv_{\LL}(u)&= \int_{\om}\LL(x) d|Du|, \label{TVLLambda}\\
\tgv_{\LL_{0}, \LL_{1}}(u)&= \min_{w\in \mathrm{BD}(\om)}\ \int_{\om}\LL_{1}(x) d|Du-w| + \int_{\om} \LL_{0}(x) d|\mathcal{E}w|. \label{TGVLLambda01}
\end{align}
The spatially dependent TV regularisation model \eqref{TVLLambda} has received significant attention in the literature, both in an infinite-dimensional and in a discrete setting,  see, e.g.,  \cite{bilevel_handbook, Pragliola_SIAMreview} and the references within. Small values of $\LL$ impose little local regularity and thus lead to better preservation of detailed parts of the image and edges. On the other hand, high values of $\LL$  impose large regularity and should thus be assigned to smooth, homogeneous areas. On the other hand, only a few works deal with spatially dependent TGV models \cite{bilevelTGV}. Here, the local interplay between $\Lambda_{0}$ and $\Lambda_{1}$ is more delicate since, additionally, these should be adjusted in a way such that reduces the staircasing effect.

\subsubsection{Conditions on the regularisation weights for well-posedness}

We now focus on the well-posedness of the problems \eqref{TVLLambda} and \eqref{TGVLLambda01}.
In order for the functional in \eqref{TVLLambda} to be well-defined, the function $\Lambda$ must be integrable with respect to the measure $Du\in \mathcal{M}(\om, \RR^{d})$, the set of $\RR^{d}$-valued finite Radon measures. Analogous integrability conditions are necessary for the functions $\Lambda_{0}$ and $\Lambda_{1}$ in order for the functional in \eqref{TGVLLambda01} to be well-defined as well. To make the corresponding minimisation problems well-posed, additional regularity is needed for these maps. For instance, existence of minimisers typically requires lower semicontinuity with respect to a suitable topology, which holds for instance when  $\Lambda$ is lower semicontinuous. Furthermore, to guarantee solutions in BV, such maps also have to be bounded away from zero. In general, the following proposition holds, which can be shown by combining results from combining \cite[Thm 3.2, Prop. 5.9]{davoli2023dyadic}, \cite[Prop. 5.1]{structuralTV} and \cite[Thm. 2.11 \& Prop. 5.17]{bredies2020higher}. 

\begin{proposition}\label{prop:existence}
Let $(\mathcal{H}, \|\cdot\|_{\mathcal{H}})$ be a Hilbert space, $f\in \mathcal{H}$ and  $\LL, \LL_{0}, \LL_{1}$ be bounded, lower semicontinuous functions which are also  bounded away from zero. Let also $A\in \mathcal{L}(L^{p}(\om), \mathcal{H})$ with $p\in (1, d^{\ast}]$,  $d^\ast=d/(d-1)$ ($d^{\ast}=\infty$ if $d=1$). Then, both problems
\begin{align}
&\min_{u\in \bv(\om)}\; \frac{1}{2}\|Au-f\|_{\mathcal{H}}^{2} +\tv_{\LL}(u), \label{weighted_TV_min}\\
&\min_{u\in \bv(\om)}\; \frac{1}{2}\|Au-f\|_{\mathcal{H}}^{2} +\tgv_{\LL_{0}, \LL_{1}}(u),  \label{weighted_TGV_min}
\end{align}
admit a solution in $\bv(\om)$.
\end{proposition}

\begin{svgraybox}
Thus, considering regularisation maps that are only lower semi-continuous and not continuous thus does not alter the well-posedness of the underlying variational problems. As we will see in the following, furthermore, this
leads to  regularisers which can be better adapted to the fine scale details of the data, eventually leading to better reconstructions than in the case of continuous regularisation maps, see Section \ref{sec:computing} for some examples.
\end{svgraybox}

\subsubsection{Continuous regularisation weights and structure of solutions}

Regarding the structure of solutions to \eqref{weighted_TV_min} and \eqref{weighted_TGV_min}, only the former has been studied in the literature, in the case of denoising and only for continuous weight functions $\LL\in \mathcal{C}(\overline{\om})$. A main result reported in \cite{jalalzai2014discontinuities} states that in the weighted TV case, the inclusion of discontinuities does not necessarily hold. In particular, is was therein shown that for $A=Id$ and $\mathcal{H}=L^{2}(\om)$, if $f\in\bv(\om)\cap L^{\infty}(\om)$, $\LL$ is Lipschitz continuous with $\nabla \LL\in \bv(\om)$, then if $u$ solves \eqref{weighted_TV_min}, then $\mathcal{H}^{d-1}(J_{u}\setminus (J_{f}\cup J_{\nabla \LL}))=0$. This means that potentially new jump discontinuities in $u$ can appear at points where the gradient of the weight $\LL$ exhibits a jump discontinuity itself.  

\begin{figure}[t]

\begin{minipage}[t]{0.49\textwidth}
\centering
\includegraphics[width=0.99\textwidth]{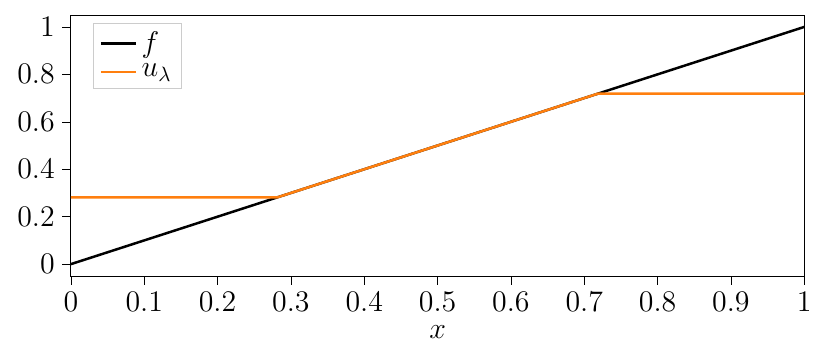}
\end{minipage}
\begin{minipage}[t]{0.49\textwidth}
\centering
\includegraphics[width=0.99\textwidth]{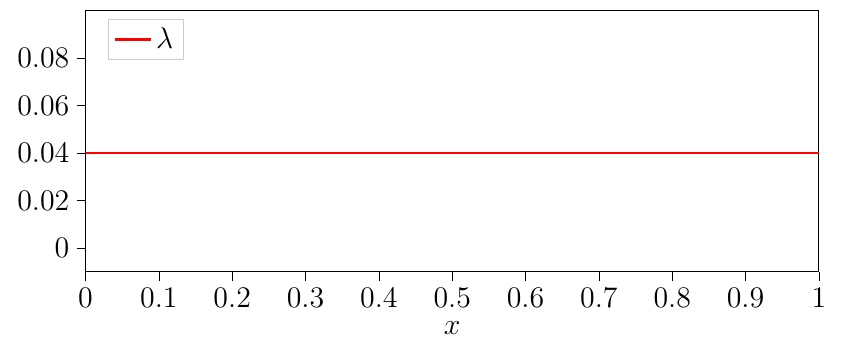}
\end{minipage}

\begin{minipage}[t]{0.49\textwidth}
\centering
\includegraphics[width=0.99\textwidth]{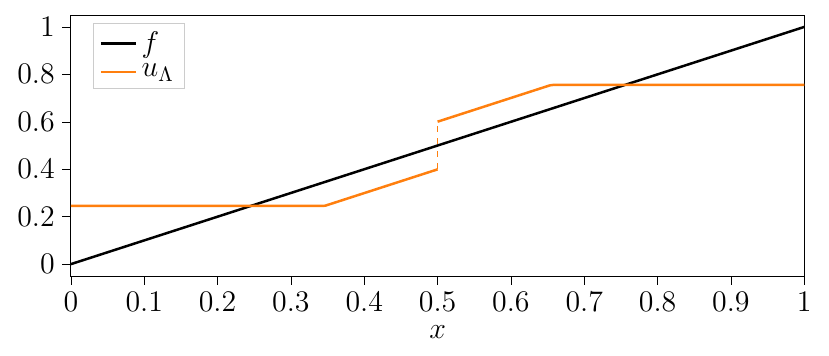}
\end{minipage}
\begin{minipage}[t]{0.49\textwidth}
\centering
\includegraphics[width=0.99\textwidth]{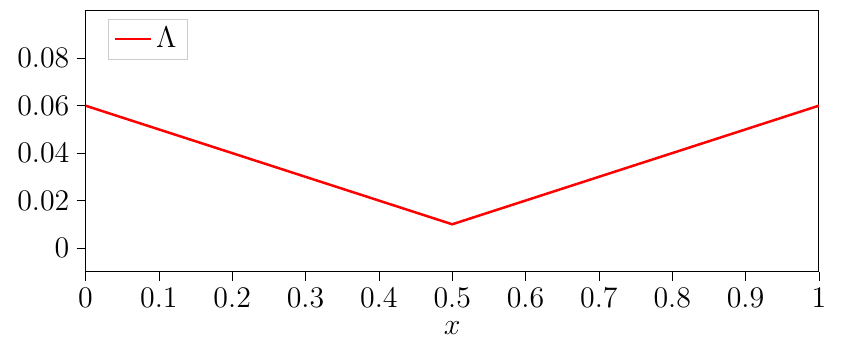}
\end{minipage}

\begin{minipage}[t]{0.49\textwidth}
\centering
\includegraphics[width=0.99\textwidth]{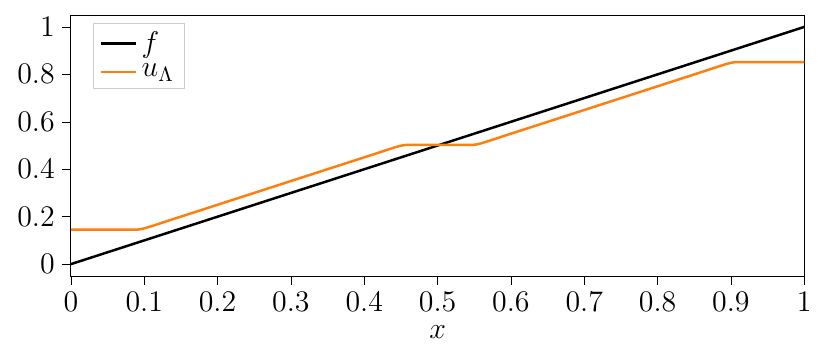}
\end{minipage}
\begin{minipage}[t]{0.49\textwidth}
\centering
\includegraphics[width=0.99\textwidth]{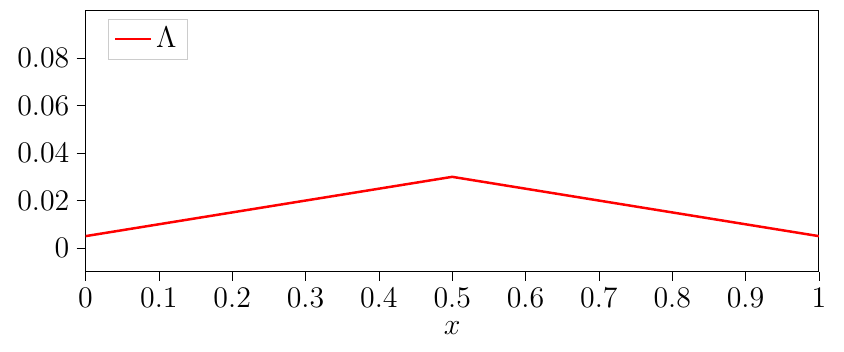}
\end{minipage}
\caption{Top: TV denoising with a scalar $\lambda$ for a simple affine function $f$. Middle: Example of creation of a new discontinuity in the solution of the  TV denoising problem with spatially varying weight $\Lambda$, at a point $x=0.5$ where $D\Lambda'(\{x\})>0$. Bottom: Creation of a flat area around the point $x=0.5$ where $D\Lambda'(\{x\})<0$.}
\label{fig:new_disc}      
\end{figure}

Again, much more can be said in dimension one: in  \cite{analyticalaspects_2017} it was shown that when $D\LL'(\{x\})>0$ and $(x-\epsilon, x+\epsilon)\subseteq \mathrm{supp}(|Du|)$ then $u$ will indeed exhibit a jump discontinuity at $x$ with the size of the jump being equal to the size of the jump of the derivative of $\LL$, i.e.\ $|Du|(\{x\})=D\LL'(\{x\})$. On the other hand, if $D\LL'(\{x\})<0$, then no new jump discontinuity is created for $u$, and in fact $u$ is constant in a neighbourhood of $x$. We show an example for each of these three cases in Fig.~\ref{fig:new_disc}. Further results from \cite{analyticalaspects_2017} include a global bound $|Du|(\om)\le |Df|(\om)$, existence in $\bv(\om)$ even for vanishing weights (provided $f\in\bv(\om)$) as well as analytic solutions for simple data and weight functions, also for non-Lipschitz continuous weights (but still in $\mathcal{C}(\overline{\om})$).

\subsubsection{Discontinuous regularisation weights pertinent to dyadic partitions}\label{sec:dyadic}
There are even more limited works that study the structure of solutions of  \eqref{weighted_TV_min} and \eqref{weighted_TGV_min} for non-continuous weights. As far as we are aware, the only relevant study -- always for denoising -- is the one of \cite{davoli2023dyadic}. One of the main results there is the following: Assume that, given a noisy $f$, we aim to approximate $u_{\mathrm{true}}$ via solutions of \eqref{weighted_TV_min} ($A=Id$) where the 
 weight $\Lambda$ is a piecewise constant function, corresponding to a dyadic partition of $\om=(0,1)^{2}$. Every value $\lambda_{L}$ of $\LL$ at a dyadic square $L$ is optimised separately, so the corresponding solution of the scalar TV problem defined on $L$ best approximates the restriction of $u_{\mathrm{true}}$ there. Then, if the weights $\lambda_{L}$ are restricted to $[c, 1/c]$ for  some $c>0$,  there exists an optimal weight $\LL$ with regards to approximating $u_{\mathrm{true}}$. In particular, after some point, further refinement of the dyadic partition is not beneficial. We note that an analogous result holds for \eqref{weighted_TGV_min} as well.

\begin{figure}[t]
\centering
\begin{minipage}[t]{0.32\textwidth}
\centering
\includegraphics[width=0.99\textwidth]{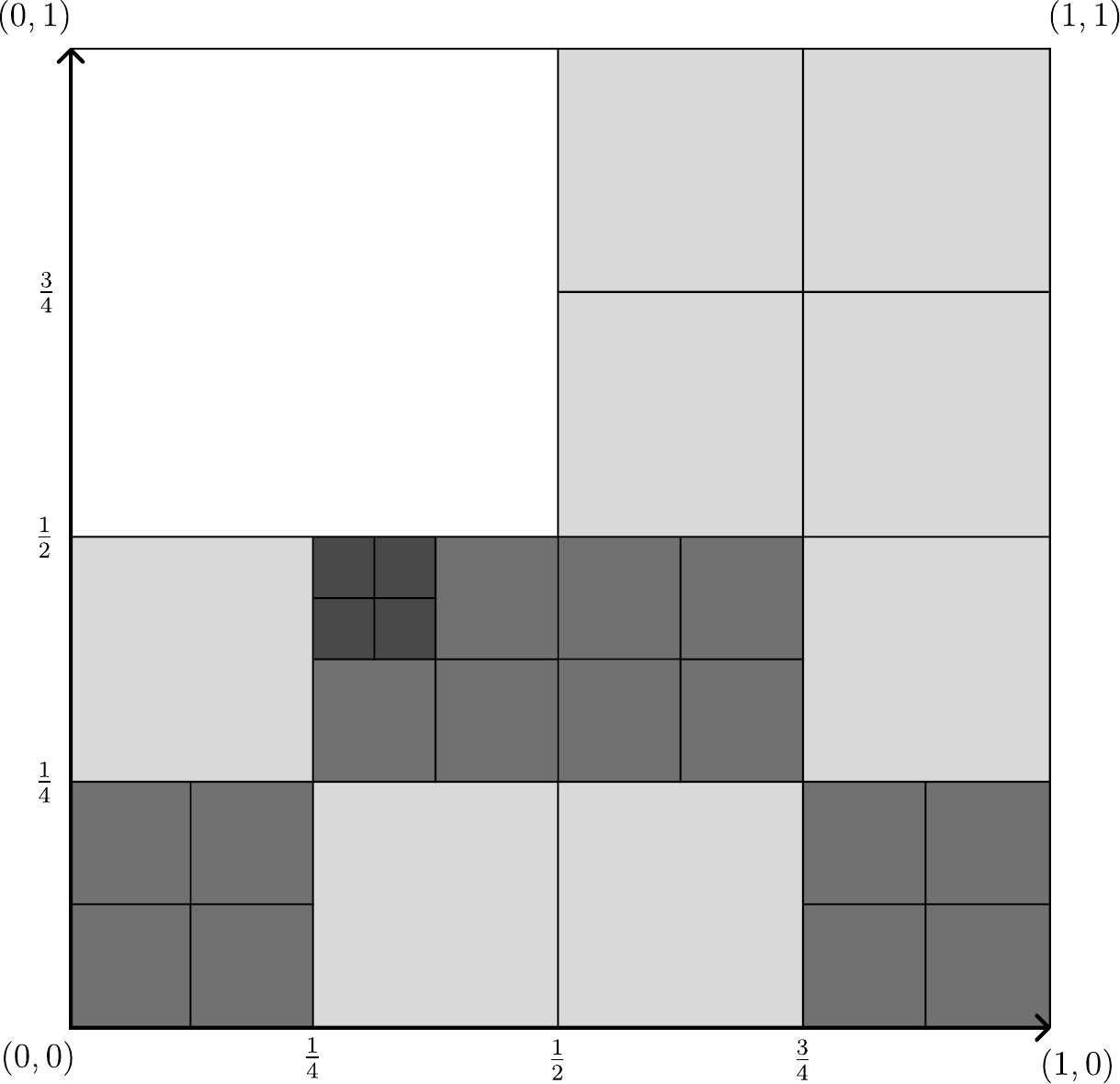}
\end{minipage}
\begin{minipage}[t]{0.32\textwidth}
\centering
\includegraphics[width=0.99\textwidth]{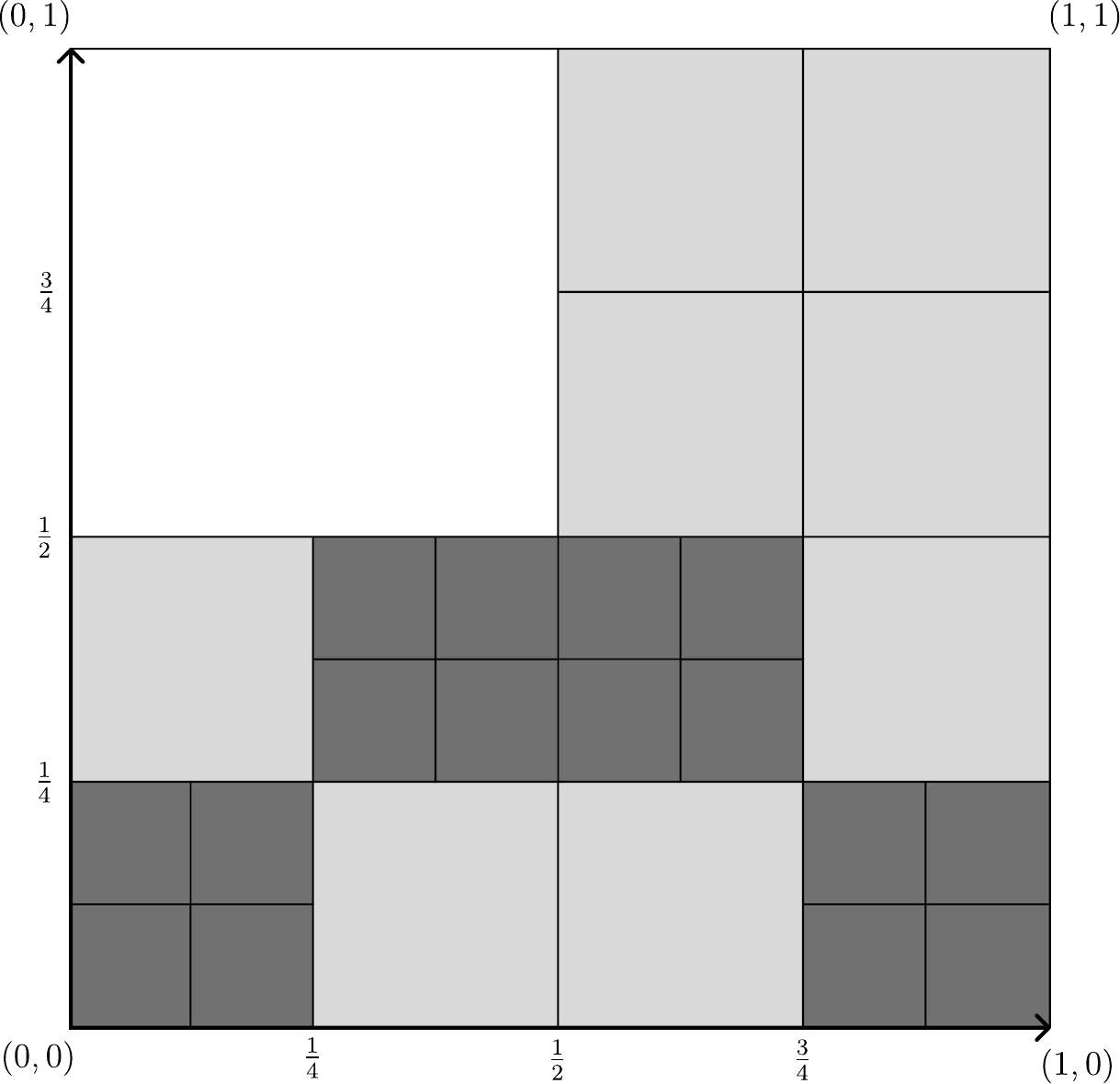}
\end{minipage}
\begin{minipage}[t]{0.32\textwidth}
\centering
\includegraphics[width=0.95\textwidth, trim={0 0 2.2cm 0},clip]{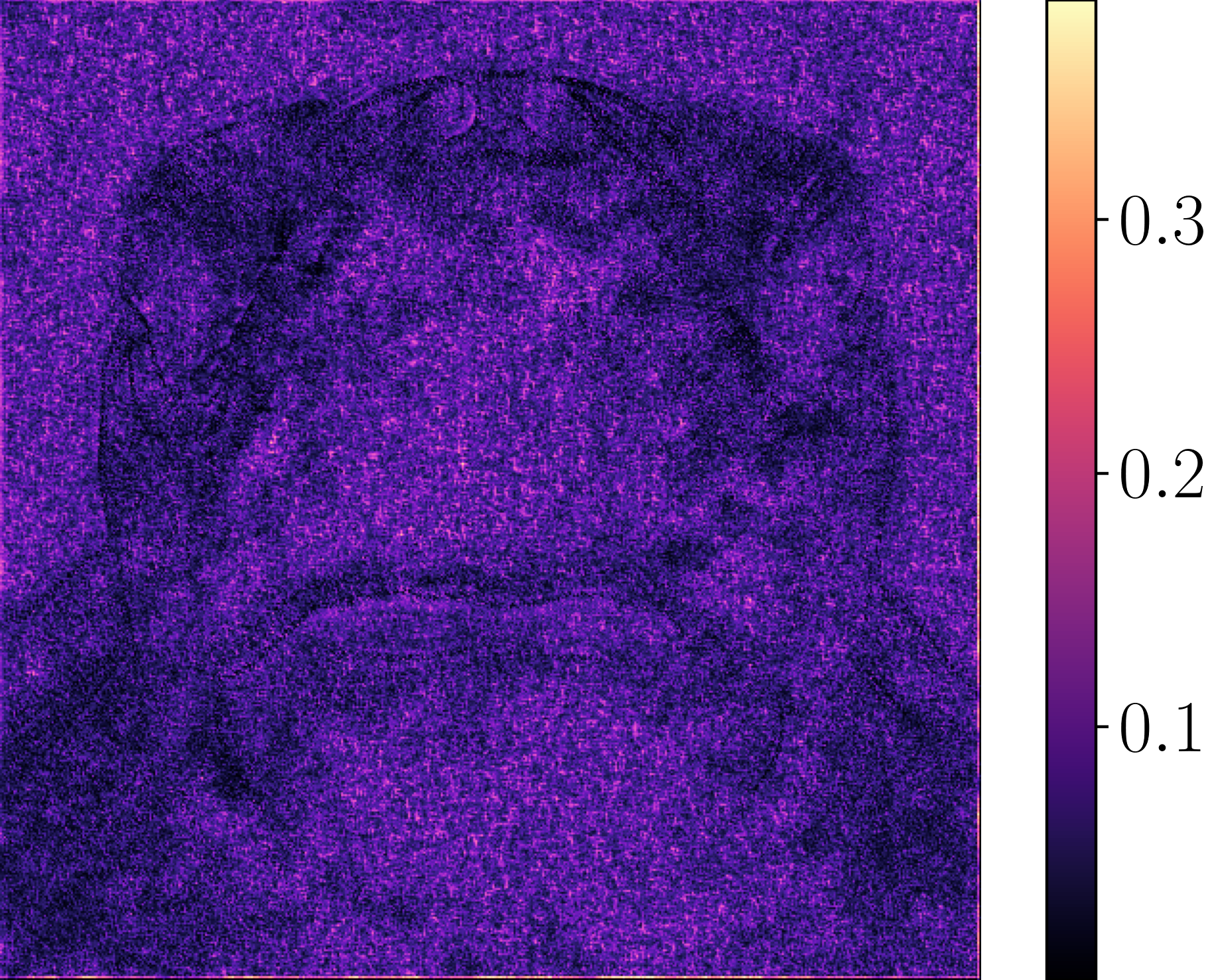}
\end{minipage}
\caption{Example of dyadic partitions of $\om=(0,1)^{2}$  considered in \cite{davoli2023dyadic} leading to discontinuous but structured regularisation weights $\LL$. For comparison, we depict again a low-regularity  weight computed by a neural network, see Section~\ref{sec:nn_unrolling} for details.}
\label{fig:dyadic_partition}      
\end{figure}

We describe this result more precisely following \cite{davoli2023dyadic}. A dyadic partition $\mathcal{L}$ of $\om=(0,1)^{2}$ consists of a division of $\om$ into squares of the type 
\[L^{\kappa}:=\left (z^{\kappa}+ \left(0,\frac{1}{2^{\kappa}}\right]^{2}\right)\cap \om,\]
where $z^{\kappa}\in \{2^{-\kappa}z\in [0,1)^{2}:z\in \mathbb{N}^{2}\}$, see Fig.~\ref{fig:dyadic_partition} for an example of such partition. The set of all dyadic partitions is denoted by $\mathcal{P}$. For every partition $\mathcal{L}$, we define
\[\tilde{\LL}_{\mathcal{L}}(x)=\sum_{L\in \mathcal{L}}\lambda_{L}\chi_{L}(x),\]
where $\chi_{L}$ is the characteristic function of the square $L$ and $\lambda_{L}$ is defined as 
\begin{equation}\label{lambdaL}
\lambda_{L}:=\inf \left \{\mathrm{argmin} \left \{ \int_{L}|u_{\mathrm{true}}-u_{\lambda, L}|\,dx: \, \lambda \in [c, 1/c] \right\} \right \},
\end{equation}
where
\begin{equation}\label{u_lambda_L}
u_{\lambda, L}:= \mathrm{argmin} \left \{ \int_{L}(u-f)^{2}dx+ TV_{\lambda}(u):\, u\in \bv(L)\right \}.
\end{equation}
and $c>0$ is some a priori fixed constant. 
Finally we set
\[\LL_{\mathcal{L}}=(\tilde{\LL}_{\mathcal{L}})^{sc^{-}}\]
that is, $\LL$ is a piecewise constant function which is the lower semicontinuous envelope of $\tilde{\LL}_{\mathcal{L}}$.  One can readily check from Prop.~\ref{prop:existence} that  the problem 
\begin{equation}\label{u_partition}
\min_{u\in \bv(\om)}\; \frac{1}{2}\int_{\om}(u-f)^{2}dx +\tv_{\LL_{\mathcal{L}}}(u),
\end{equation}
has a unique solution $u_{\mathcal{L}}$ that depends only on the partition $\mathcal{L}$. Then, one of the main results of \cite{davoli2023dyadic} states that the problem
\begin{equation}\tag{$P$}\label{min_over_partitions}
\min\left\{\int_{\om} (u_{\mathrm{true}}-u_{\mathcal{L}})^{2}\, dx: \, \mathcal{L}\in \mathcal{P} \right \},
\end{equation} 
has a solution. This means that for these kinds of weights, there exists a minimum square length, after which it is not beneficial to refine the partition further in order to produce a weight $\LL$ that leads to a better approximation of $u_{\mathrm{true}}$. We note that the condition $\lambda\in [c,1/c]$ in  \eqref{lambdaL} is crucial for this existence. 

Note that the problem \eqref{min_over_partitions} is different than the following modified version:
\begin{equation}\tag{$\hat{P}$}\label{min_over_partitions_full}
\left \{
\begin{aligned}
&\inf\left\{\int_{\om} (u_{\mathrm{true}}-u_{\LL})^{2}\, dx: \,\LL=\left (\sum_{L\in \mathcal{L}}\lambda_{L}\chi_{L}\right)^{sc^{-}}, \, \lambda_{L}\in [c,1/c], \,  \mathcal{L}\in \mathcal{P}\right \},\\[0.5em]
&\text{such that}\quad  u_{\LL}=\underset{u\in\bv(\om)}{\mathrm{argmin}}\; \frac{1}{2}\int_{\om}(u-f)^{2}dx +\tv_{\LL}(u).
\end{aligned}\right.
\end{equation} 
Indeed the determination of the scalar weights $\lambda_{L}$ in \eqref{min_over_partitions} is done based on the local fitting of a TV minimiser on the square $L$ only, whereas in \eqref{min_over_partitions_full} these are chosen based on the denoising result on the whole domain via the full weight $\Lambda$. Obviously, if $u$ and $\hat{u}$ are minimisers of  \eqref{min_over_partitions} and  \eqref{min_over_partitions_full} respectively it follows that
\begin{equation}\label{full_better}
\int_{\om} (u_{\mathrm{true}}-\hat{u})^{2}\, dx\le \int_{\om} (u_{\mathrm{true}}-u)^{2}\, dx.
\end{equation}
In order to see why the strategy for the selection of $\lambda_{L}$ in \eqref{min_over_partitions} might be suboptimal, which would happen in the case of a strict inequality in \eqref{full_better}, consider the following example in dimension one.
Let $f:(0,1)\to \RR$, with $f(x)=\chi_{(1/2,1)}(x)+ h\chi_{(1/2+\delta, 1)}(x)$, for some $h>0$ and $\delta\in (0,1/2)$. The idea is to define $u_{\mathrm{true}}$ in a way such that $u_{\Lambda}=u_{\mathrm{true}}$ for some lower semicontinuous piecewise constant $\LL=\lambda_{1}\chi_{(0,1/2]}+\lambda_{2}\chi_{(1/2,1)}$, and thus $u_{\LL}=\hat{u}$ the minimiser of \eqref{min_over_partitions_full} with zero objective, but on the other hand it holds $u\ne u_{\mathrm{true}}$ for the minimiser $u$ of \eqref{min_over_partitions}.

\begin{figure}[t]
\centering
\begin{minipage}[t]{0.49\textwidth}
\centering
\includegraphics[width=0.99\textwidth]{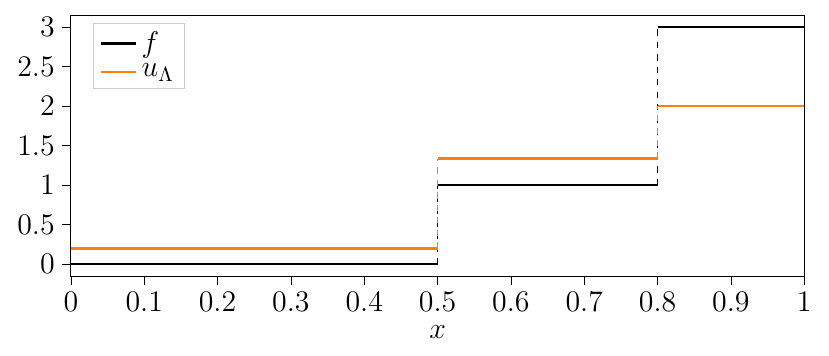}
\end{minipage}
\begin{minipage}[t]{0.49\textwidth}
\centering
\includegraphics[width=0.99\textwidth]{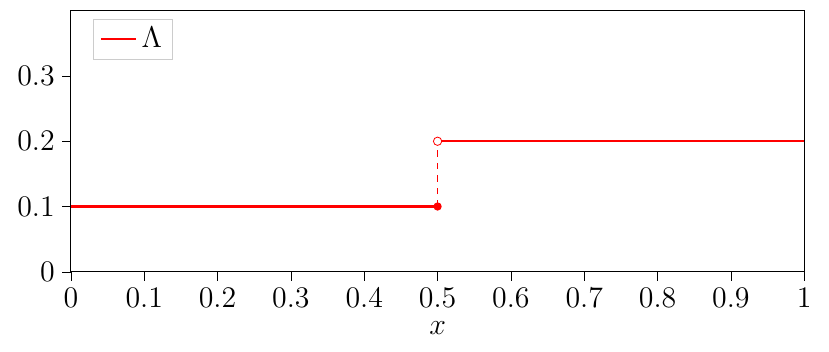}
\end{minipage}
\caption{Example of a data function $f$ and $u_{\mathrm{true}}:=u_{\LL}$, which is solution of \eqref{min_over_partitions_full} with zero objective but cannot be a solution of the problem \eqref{min_over_partitions}.}
\label{fig:counterxample}      
\end{figure}

Adapting the work in \cite{analyticalaspects_2017} for a weight like  the $\LL$ above, we can easily compute $u_{\LL}$ for the different values of $\lambda_{1}, \lambda_{2}$. For instance 
if 
\[\lambda_{1}<\lambda_{2} \text{ and } \lambda_{2}<2h\delta\left(\tfrac{1}{2}-\delta\right)+2\left(\tfrac{1}{2}-\delta\right)\lambda_{1},\]
then
\begin{equation}\label{counterex_f_2}
u_{\Lambda}(x)=
\begin{cases}
2\lambda_{1} & \text{ if }x\in \left(0,\tfrac{1}{2}\right],\\
1+\frac{\lambda_{2}-\lambda_{1}}{\delta} & \text{ if }x\in \left(\tfrac{1}{2},\tfrac{1}{2}+\delta\right],\\
1+h-\frac{\lambda_{2}}{1-\frac{1}{2}-\delta} & \text{ if }x\in \left(\tfrac{1}{2}+\delta,1\right).
\end{cases}
\end{equation}
In Fig.~\ref{fig:counterxample} we provide an example for $\lambda_{1}=0.1$, $\lambda_{2}=0.2$, $h=2$, and $\delta=0.3$  and we set $u_{\mathrm{true}}=u_{\LL}$ for such $\LL$. 

It now suffices to show that the previous function of the type \eqref{counterex_f_2} cannot be a solution of \eqref{min_over_partitions}. Indeed, on one hand, one can easily compute the set of TV solutions $u_{\lambda}$ for scalar $\lambda>0$ (which can be seen as a dyadic weight $\LL$ pertinent to the trivial partition $\mathcal{L}$=\{(0,1)\})  and confirm that solutions where $u_{\lambda}>f$ both in $(0,1/2)$ and $(1/2, 1/2+\delta)$ cannot occur. On the other hand, any other partition $\mathcal{L}$ would lead to $\lambda_{L}=c$ for any segment $L\subseteq (0,1/2)$ since $f$ is constant there. Thus, in view of \eqref{counterex_f_2}, it can be checked that if $c$ is chosen such that $c<2\lambda_{1}$, $u_{\mathcal{L}}$ cannot be equal to $u_{\mathrm{true}}$ in $(0,1/2)$.

Of course, in general it is an open question which conditions guarantee the existence of solutions for the problem \eqref{min_over_partitions_full}. This may involve some compactness conditions for the space of admissible dyadic weights $\Lambda$. For a discussion on restricting to general lower semicontinuous weights, and related existence problems, we refer to \cite{pagliari2022bilevel}.

\section{Discretisation and Bayesian interpretation}

In this section, we discuss the discretisation and the Bayesian interpretation of the weighted regularisation models, also briefly mentioning how the choice of the regularisation weights can be seen in this context.

\subsection{Discretisation}

We briefly describe the discretisation of the differential operators entering
the weighted TV and TGV models. Let $u\in\mathbb{R}^{N}$ denote the vectorised image
defined on a Cartesian grid, and let
$
    D=(D_h,D_v)
$
denote the discrete gradient, where $D_h,D_v\in\mathbb{R}^{N\times N}$
are first-order finite-difference operators in the horizontal and vertical directions, respectively.
For a discrete vector field $w=(w_1,w_2)$, the symmetric
gradient is correspondingly discretised as
\begin{equation}
    Ew=
    \begin{pmatrix}
        D_hw_1 &
        \frac{1}{2}(D_v w_1+D_h w_2)
        \\[0.2em]
        \frac{1}{2}(D_v w_1+D_h w_2) &
        D_v w_2
    \end{pmatrix}.
    \label{eq:discrete_symgrad}
\end{equation}

For $i=1,\ldots,N$, denoting by $\Lambda_i>0$ and $\Lambda_{\ell,i}>0$, $\ell=0,1$, the pixelwise
values of the spatially varying weights, the discrete counterparts of the
weighted TV and weighted TGV functionals, \eqref{TVLLambda} and \eqref{TGVLLambda01}, read
\begin{align}
    \tv_{\Lambda}^{h}(u)
    &=
    \sum_{i=1}^{N}
    \Lambda_i |(Du)_i|_2,
    \label{eq:discrete_wtv}
    \\
    \tgv_{\Lambda_0,\Lambda_1}^{h}(u)
    &=
    \min_{w}
    \left\{
    \sum_{i=1}^{N}
    \Lambda_{1,i}|(Du-w)_i|_2
    +
    \sum_{i=1}^{N}
    \Lambda_{0,i}|(Ew)_i|_F
    \right\},
    \label{eq:discrete_wtgv}
\end{align}
where $|\cdot|_2$ and $|\cdot|_F$ denote the Euclidean and Frobenius norms,
respectively.

\subsection{Bayesian interpretation} \label{sec:discrete_bayesian}

Having introduced the discretised quantities as above, we report in the following how the reconstruction model \eqref{general_min_L2} can be derived from a Bayesian perspective and discuss the prior induced by the spatially varying weights. Before that, we recall the likelihood term corresponding to the Gaussian noise model considered. 
Denoting by $A\in\mathbb{R}^{K\times N}$ the discretised forward operator, by $f\in\mathbb{R}^{K}$ the measurement vector and with $\eta\sim\mathcal{N}(0,\sigma^{2}I_K)$ an i.i.d.\ Gaussian noise vector
with variance $\sigma^{2}>0$, we have that the likelihood function reads
\[
    \pi(f\mid u)
    =
    (2\pi\sigma^{2})^{-K/2}
    \exp\left(
        -\frac{1}{2\sigma^{2}}\|Au-f\|_2^{2}
    \right).
\]
Therefore
\[
    -\log\pi(f\mid u)
    =
    \frac{1}{2\sigma^{2}}\|Au-f\|_2^{2}
    +
    \frac{K}{2}\log(2\pi\sigma^{2}),
\]
where, for fixed $\sigma$, the last term does not depend on $u$ and can be
neglected upon optimisation. Combining such Gaussian likelihood with the
prior $\pi(u\mid\Lambda)$ derived below via Bayes' rule,
$
    \pi(u\mid f,\Lambda)
    \propto
    \pi(f\mid u)\,\pi(u\mid\Lambda),
$
and maximising the posterior $\pi(u\mid f,\Lambda)$ with
respect to $u$ (that is, MAP estimation), turns out to be equivalent to minimising
$
    \frac{1}{2\sigma^{2}}\|Au-f\|_2^{2} - \log\pi(u\mid\Lambda).
$
Here, the 
models for
$- \log\pi(u\mid\Lambda)$ depending on the spatially adaptive weights can be defined following the non-stationary Markov random field viewpoint described in
\cite{Pragliola_SIAMreview}. 

The weighted TV Gibbs prior can be motivated by assigning a local exponential (half-Laplacian) potential to each
gradient magnitude
$
    z_i(u):=|(Du)_i|_2,\;i=1,\ldots,N
$
with spatially varying rate $\Lambda_i>0$. Formally, taking the product of these local densities gives
\begin{equation}
    \widetilde{\pi}(u\mid\Lambda)
    =
    \left(\prod_{i=1}^{N}\Lambda_i\right)
    \exp\left\{
        -\sum_{i=1}^{N}
        \Lambda_i |(Du)_i|_2
    \right\}.
    \label{eq:bayes_wtv}
\end{equation}
Since neighbouring discrete gradients generally involve common pixel
values, the quantities $z_i(u)$ should not be interpreted as 
independent random variables under the resulting image prior. Rather, the
product construction provides local Gibbs potentials which, after
normalisation,
\[
   \pi(u\mid\Lambda)
    =
    c(\Lambda)\,\widetilde{\pi}(u\mid\Lambda),
\]
define a non-stationary Markov random field on the image. 
Consequently,
\begin{equation}
    -\log\pi(u\mid\Lambda)
    =
    \tv_{\Lambda}^{h}(u)
    -\sum_{i=1}^{N}\log\Lambda_i
    -\log c(\Lambda),
    \label{eq:bayes_wtv_log}
\end{equation}
where, for fixed $\Lambda$, the last two terms do not depend on $u$ and can
therefore be neglected upon optimisation.  In this
interpretation, large values of $\Lambda_i$ concentrate the local gradient
distribution near zero and impose stronger regularisation, whereas small
values allow larger gradients and thus favour the preservation of edges and
fine structures.

A similar interpretation can be associated with the weighted TGV regularisation by retaining
the auxiliary vector field $w$ as a latent variable. In analogy with the
construction above, we introduce the first- and second-order term
magnitudes
\[
    z_{1,i}(u,w):=|(Du-w)_i|_2,
    \qquad
    z_{0,i}(w):=|(Ew)_i|_F.
\]
Motivated by the exponential (half-Laplacian) modelling of local
gradient magnitudes used for weighted TV, we associate with the
first- and second-order terms the local Gibbs potentials
\[
    V_{1,i}(u,w;\Lambda_{1,i})
    :=
    \Lambda_{1,i}|(Du-w)_i|_2,
    \qquad
    V_{0,i}(w;\Lambda_{0,i})
    :=
    \Lambda_{0,i}|(Ew)_i|_F,
\]
As for the weighted TV, notice that these local
quantities are not assumed to be statistically independent.
The resulting non-stationary Gibbs prior on the augmented variable
$(u,w)$ can thus be formally written as
\begin{equation}
    \pi(u,w\mid\Lambda_0,\Lambda_1)
    :=
   c(\Lambda_0,\Lambda_1)
    \exp\left\{
        -\sum_{i=1}^{N}
        \Lambda_{1,i}|(Du-w)_i|_2
        -
        \sum_{i=1}^{N}
        \Lambda_{0,i}|(Ew)_i|_F
    \right\},
    \label{eq:bayes_wtgv}
\end{equation}
where $c(\Lambda_0,\Lambda_1)>0$ denotes the corresponding normalisation
constant, whenever finite. Its negative
logarithm reads
\begin{align}
    -\log\pi(u,w\mid\Lambda_0,\Lambda_1)
    ={}&
    \sum_{i=1}^{N}
    \Lambda_{1,i}|(Du-w)_i|_2
    +
    \sum_{i=1}^{N}
    \Lambda_{0,i}|(Ew)_i|_F
    \nonumber\\
    &-\log c(\Lambda_0,\Lambda_1).
    \label{eq:bayes_wtgv_log}
\end{align}
For fixed weights, the normalisation term is independent of $(u,w)$ and
can therefore be neglected in MAP estimation. The $(u,w)$-dependent
part of the negative log-prior is thus exactly the weighted TGV energy
before minimisation with respect to the auxiliary field $w$.

\begin{svgraybox}
The parameter maps $\Lambda$ above have so far been regarded as
fixed and strictly positive. A natural extension is to treat them as additional
unknowns within a \emph{hierarchical} Bayesian model, that is, by assigning suitable
hyperpriors to their components with or without additional spatial regularity assumptions. This viewpoint has been extensively explored
in Bayesian inverse problems, see, e.g.~\cite{CalvettiPragliolaSomersalo2020,CalvettiPragliolaSomersaloStrang2020}, where gamma and
generalised-gamma hyperpriors are used to model unknown local prior parameters
and estimate them jointly with the quantity of interest. 
\end{svgraybox}

\section{Practical computation of regularisation weights across regularity classes}
\label{sec:computing}
We now turn our attention to more practical aspects such as how to choose and impose regularity on these weights at the discrete setting in practice.

\subsection{Bilevel optimisation for learning regularisation weights }

A large family of approaches for determining regularisation weights is based on bilevel optimisation, a first instance of which was already presented for problems \eqref{min_over_partitions_full} and \eqref{min_over_partitions}. In general, one aims to solve a problem of the type
\begin{equation}\label{bilevel}
\left \{
\begin{aligned}
&\inf_{\LL\in \mathcal{A}_{adm}}\; \mathcal{F}(u_{\LL}, \LL),\\[0.5em]
&\text{such that}\quad  u_{\LL}\in\underset{u\in \RR^{N}}{\mathrm{argmin}}\; \frac{1}{2}\|Au-f\|_{2}^{2} +\mathcal{R}(u;\mathrm{\LL}),
\end{aligned}\right.
\end{equation} 
for a suitable upper level objective $\mathcal{F}$.
Depending on whether $\mathcal{F}$ is defined in terms of features coming from the (typically unknown) ground truth image $u_{\mathrm{true}}$ or not, these approaches are separated into supervised and self-supervised ones. Usually, self-supervised approaches involve a statistics-based upper level objective which tries to enforce the residual $Au-f$ to have characteristics of Gaussian noise (whiteness-principle), \cite{Fehrenbach_2015, hintermuller2017optimal, santambrogio2024whiteness,PragliolaWTV_SSM2025}. There is a vast literature on the topic of bilevel optimisation for determining these weights, and we refer the interested reader to the 
the reviews  and the references therein \cite{Calatroni_bilevellearning, Crockett_2022, bilevel_handbook, Reyes2023}. 

\begin{figure}[t]
\centering
\begin{minipage}[t]{0.24\textwidth}
\centering
       \begin{tikzpicture}[spy using outlines={rectangle, white, magnification=2, size=0.9cm, connect spies}]
  	  \node[anchor=south west,inner sep=0]  at (0,0) {
        \includegraphics[height=2.9cm]{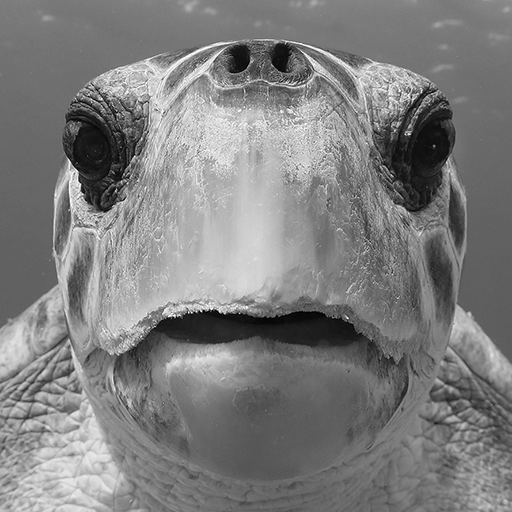}};
   		 \spy on (0.5, 2.2) in node [left] at (2.75, 0.5);
  	  \end{tikzpicture}
\end{minipage}\hspace{1.5em}
\begin{minipage}[t]{0.3\textwidth}
       \begin{tikzpicture}[spy using outlines={rectangle, white, magnification=2, size=0.9cm, connect spies}]
  	  \node[anchor=south west,inner sep=0]  at (0,0) {
\includegraphics[height=2.9cm]{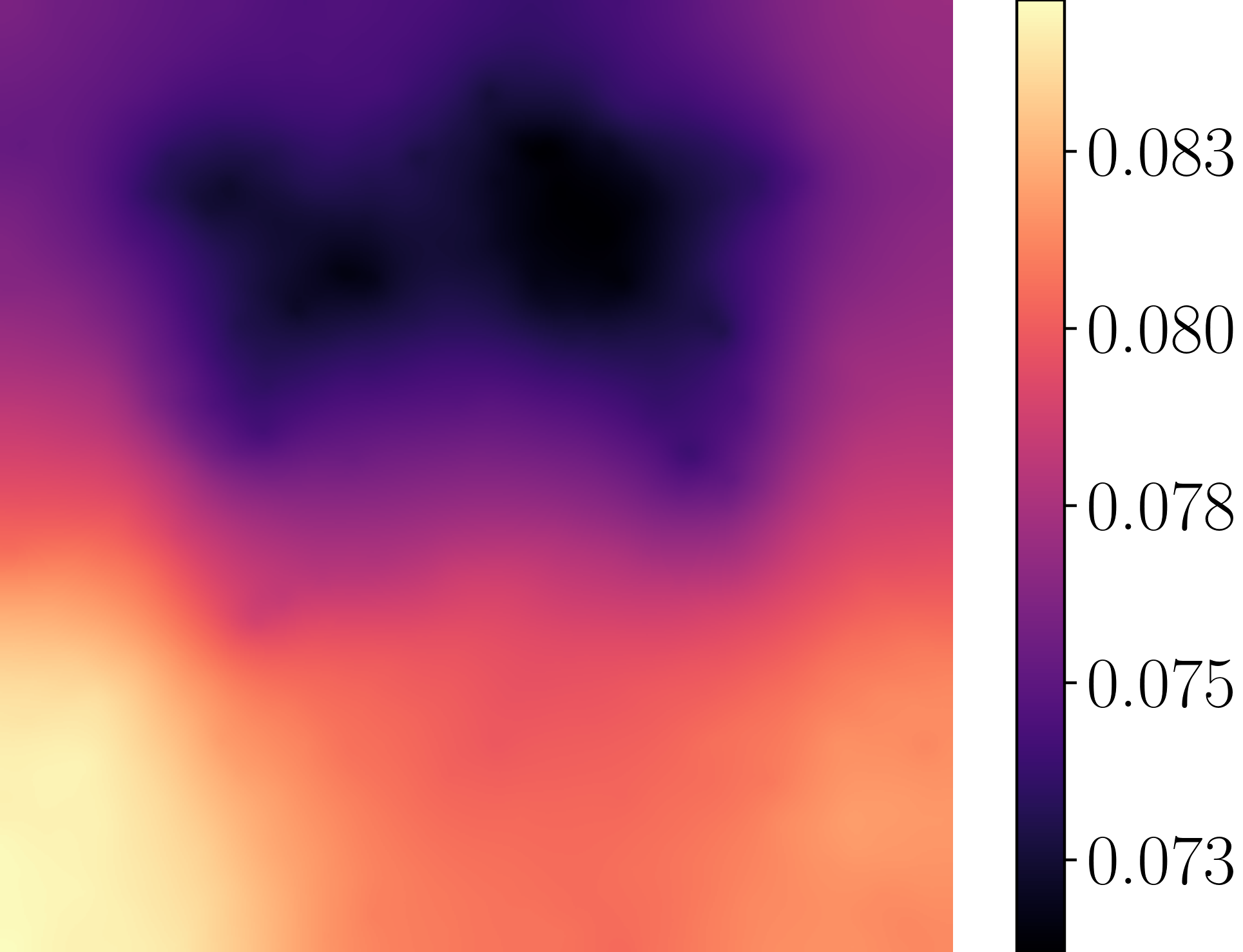}};
   		 \spy on (0.5, 2.2) in node [left] at (2.75, 0.5);
  	  \end{tikzpicture}
\end{minipage}\hspace{0.75em}
\begin{minipage}[t]{0.3\textwidth}
\begin{tikzpicture}[spy using outlines={rectangle, white, magnification=2, size=0.9cm, connect spies}]
  	  \node[anchor=south west,inner sep=0]  at (0,0) {
\includegraphics[height=2.9cm]{figures/results/denoising/turtle/turtle_denoised_UTV_Lambda_2026_08_24_enhanced}};
   		 \spy on (0.5, 2.2)  in node [left] at (2.75, 0.5);
  \end{tikzpicture}
\end{minipage}

\begin{minipage}[t]{0.24\textwidth}
\centering
{\footnotesize Clean image}
\end{minipage}\hspace{1.5em}
\begin{minipage}[t]{0.3\textwidth}
{\footnotesize  $\LL$ of high regularity \cite{hintermuller2017optimal}}
\end{minipage}\hspace{0.75em}
\begin{minipage}[t]{0.3\textwidth}
{\footnotesize  $\LL$ of low regularity  \cite{kofler2023learning}}
\end{minipage}%
\vspace{0.5em}

\begin{minipage}[t]{0.24\textwidth}
\centering
\phantom{clean}
\end{minipage}\hspace{1.5em}
\begin{minipage}[t]{0.3\textwidth}
{\hspace{1.45cm}\Large$\downarrow$}
\end{minipage}\hspace{0.75em}
\begin{minipage}[t]{0.3\textwidth}
{\hspace{1.45cm}\Large$\downarrow$}
\end{minipage}
\vspace{0.5em}

\begin{minipage}[t]{0.24\textwidth}
\centering
\begin{tikzpicture}[spy using outlines={rectangle, white, magnification=2, size=0.9cm, connect spies}]
  	  \node[anchor=south west,inner sep=0]  at (0,0) {
\includegraphics[height=2.9cm]{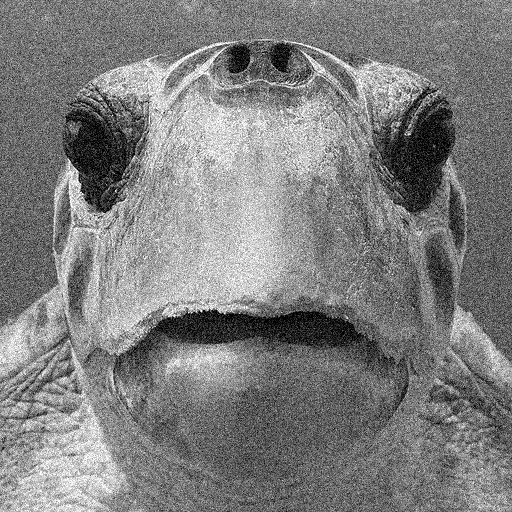}};
   		 \spy on (0.5, 2.2) in node [left] at (2.75, 0.5);
  \end{tikzpicture}
\end{minipage}\hspace{1.5em}
\begin{minipage}[t]{0.3\textwidth}
       \begin{tikzpicture}[spy using outlines={rectangle, white, magnification=2, size=0.9cm, connect spies}]
  	  \node[anchor=south west,inner sep=0]  at (0,0) {
        \includegraphics[height=2.9cm]{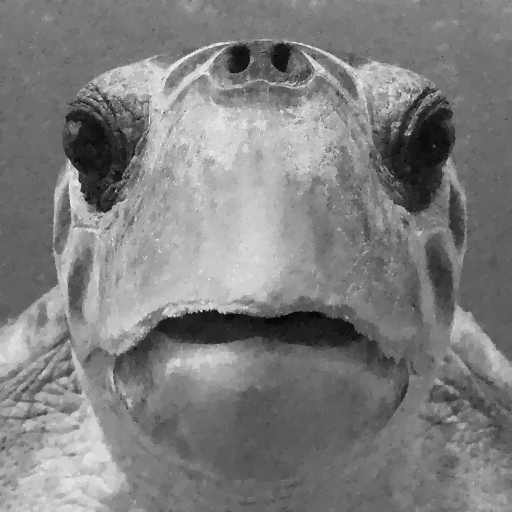}};
   		 \spy on (0.5, 2.2) in node [left] at (2.75, 0.5);
  	  \end{tikzpicture}
\end{minipage}\hspace{0.75em}
\begin{minipage}[t]{0.3\textwidth}
       \begin{tikzpicture}[spy using outlines={rectangle, white, magnification=2, size=0.9cm, connect spies}]
  	  \node[anchor=south west,inner sep=0]  at (0,0) {
       \includegraphics[height=2.9cm]{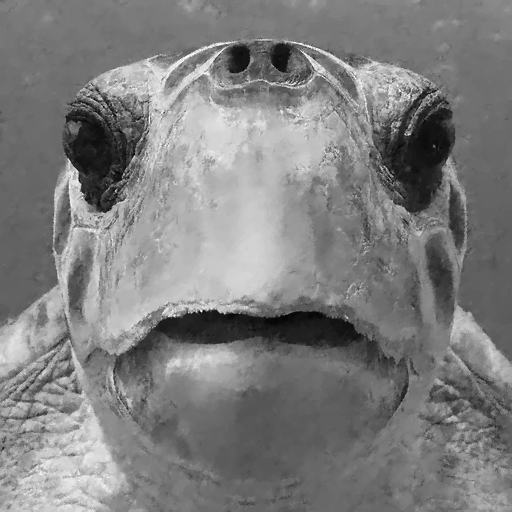}};
   		 \spy on (0.5, 2.2) in node [left] at (2.75, 0.5);
  	  \end{tikzpicture}
\end{minipage}%

\begin{minipage}[t]{0.24\textwidth}
\centering
{\footnotesize Noisy image}
\end{minipage}\hspace{1.5em}
\begin{minipage}[t]{0.3\textwidth}
{\footnotesize TV denoising with $\LL$ of high regularity. PSNR: 27.42}
\end{minipage}\hspace{0.75em}
\begin{minipage}[t]{0.3\textwidth}
{\footnotesize TV denoising with $\LL$ of low regularity. PSNR: 27.92}
\end{minipage}%
\caption{TV denoising with spatially varying regularisation weights $\LL$ of different regularity. Note that in the discrete setting,  weights of high regularity are understood as the discrete analogues of continuous weights, with smooth transitions between the weight values; see \cite{hintermuller2017optimal}. On the other hand, weights of low regularity are understood as discrete analogues of discontinuous weights, with abrupt transitions. The latter ones can better adapt to the image details, leading to better reconstructions as the images above showcase. Note the different scale between the two weights. }
\label{fig:denoising_regularity}      
\end{figure}

\begin{svgraybox}
The desired regularity of $\LL$ can be imposed by defining a suitable term in $\mathcal{F}$. For instance, the approaches outlined in \cite{bilevel_handbook} employ a $H^{1}$-norm on $\Lambda$ which, for $d=2$ and in the continuous setting, together with certain regularity results, guarantees that the resulting weights are continuous. An illustration of these weights in the discrete setting is reported in Fig.~\ref{fig:denoising_regularity}.
\end{svgraybox}

Despite their popularity and their rich mathematical theory, bilevel approaches present severe computational bottlenecks. Due to their nested structure, vanilla approaches require the lower-level reconstruction problem to be solved many times during the numerical implementation and separately for every image, which is computationally intense. 
Thus, in the literature, it is mainly applied for denoising and simple 2D tasks, much less for more advanced and larger-scale problems. Furthermore, there is a limit to the quality of the regularisation weights --and, as a consequence, to the quality of the corresponding reconstructions as well-- that can result from handcrafted statistics-based upper-level objectives. In the following, we present a recent approach that employs the power of deep neural networks to estimate regularisation parameters.

\subsection{Neural network-unrolled algorithmic based-approach for learning regularisation weights }\label{sec:nn_unrolling}
Neural networks and deep learning have had a tremendous influence in imaging and inverse problems over the last decade. While in general their use leads to superior performance in comparison to the aforementioned model-based approaches, there have been concerns about their black-box nature, instabilities and limited interpretabilty \cite{Antun_2020}. Indeed, little can be said about the structure of image reconstructions that are output of deep neural networks, which, as we saw in the previous sections, is in strong contrast to the TV and TGV ones. This has led to the development of hybrid model-based and data-driven approaches, which combine the interpretability and reconstruction guarantees of model-based methods with the versatility and superior performance of deep neural networks. The literature even on this subfield of deep learning is already vast and many review papers aim to cover the continuous progress of this area, e.g. \cite{arridge2019solving, habring2024neural, HERTRICH202685,   kamilov2023plug, Kofler2024, monga2021algorithm, mukherjee2023learned}.

Here, we  review a particular approach, originating from \cite{kofler2023learning}, that falls within this hybrid category and operates at the level of the regularisation parameters, rather than the image itself. 
\begin{figure}[!t]
\centering
\begin{minipage}[t]{\textwidth}
\centering
\includegraphics[width=0.99\textwidth]{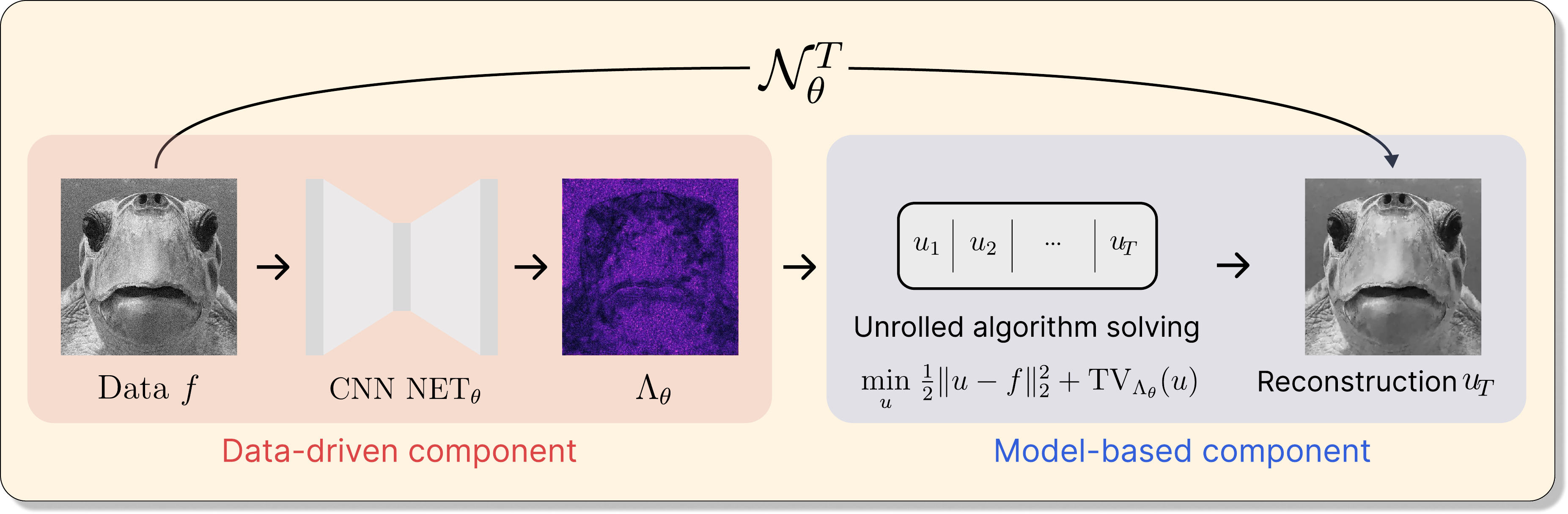}
\end{minipage}
\caption{Visualisation of the unrolled network $\mathcal{N}_{\theta}^{T}: f\mapsto \LL_{\theta}\mapsto$ reconstruction $u_{T}$ which corresponds to a solution of a model-based reconstruction (here shown for TV denoising) with a neural network-inferred spatially varying regularisation parameter $\LL_{\theta}$.}
\label{fig:unrolled}      
\end{figure}
Let $u_{n}:=S^{n}(u_{0}, f, \LL, A)$ denote the iterates of some algorithm solving \eqref{weighted_TV_min} or \eqref{weighted_TGV_min}, where  $u_{0}$ is some suitable initialisation  e.g. $u_{0}=A^{\ast}f$ with $A^{\ast}$ being the adjoint of $A$. This means that $u_{n}\to u_{\Lambda}$ for some solution of \eqref{weighted_TV_min} or \eqref{weighted_TGV_min}. Next we denote with $\mathrm{NET}_{\theta}:A^{\ast}f \mapsto \LL_{\theta}$ a deep convolutional neural network with learnable parameters $\theta$, e.g.\ a U-Net \cite{Ronneberger2015}. Then,  for a \emph{fixed} $T\in \mathbb{N}$, we define the overall network
\begin{equation}\label{unrolled_network}
\mathcal{N}_{\theta}^{T}(f)= S^{T}(A^{\ast}(f), f, \mathrm{NET}_{\theta}(A^{\ast}f), A).
\end{equation}
The unrolled network \eqref{unrolled_network} can then be trained end-to-end in a supervised fashion using a dataset of data-ground truth pairs $(f^{i}, u_{\mathrm{true}}^{i})_{i=1}^{M}$, and an appropriate pairwise distance function $l$, 
\begin{equation}\label{supervised_learning}
\min_{\theta}\; \mathrm{Loss}(\theta):= \frac{1}{M} \sum_{i=1}^{M} l(\mathcal{N}_{\theta}^{T}(f^{i}), u_{\mathrm{true}}^{i}).
\end{equation}
Notably, the training data do not have to be pairs of ground truths and optimal regularisation parameters, as is done e.g. in \cite{Afkham_2021}.
 In other words, during training, the parameters $\theta$ of the deep neural network $\mathrm{NET}_{\theta}$ are optimised in such a way that the output of the network is a spatially varying parameter $\Lambda_{\theta}$  for which the $T$-th iterate is as close as possible to the ground truth.
This approach is outlined in Fig.~\ref{fig:unrolled} for the special case of denoising and spatially varying TV. All the black-box nature of the deep neural network operates at the level of the regularisation parameter, while the reconstruction remains model-based and interpretable as a solution of a variational problem with a classical regulariser. This approach was introduced in \cite{kofler2023learning} for the case of TV and applied to a variety of inverse imaging problems like dynamic MRI, dynamic denoising, quantitative MRI and computerised tomography. We note that for dynamic problems, where in addition to the two spatial variables, there is also a time variable, the network outputs regularisation parameters that are spatio-temporally dependent. The approach has been subsequently extended to TGV  \cite{Wu_2025} and to convolutional synthesis regularisation \cite{Kofler_2025, schulz2026learning}. We refer to these papers for specific details and extensive comparisons. In short, this approach consistently produces spatio-(temporal) varying regularisation parameters that lead to better reconstructions compared to the scalar versions and are sometimes on par with state-of-the-art pure deep learning approaches. 

\subsection{Numerical experiments in image denoising and MRI reconstruction}

\begin{figure}[!ht]
\centering
\begin{minipage}[t]{0.19\textwidth}
       \begin{tikzpicture}[spy using outlines={rectangle, white, magnification=2, size=0.7cm, connect spies}]
  	  \node[anchor=south west,inner sep=0]  at (0,0) {
\includegraphics[width=0.99\textwidth]{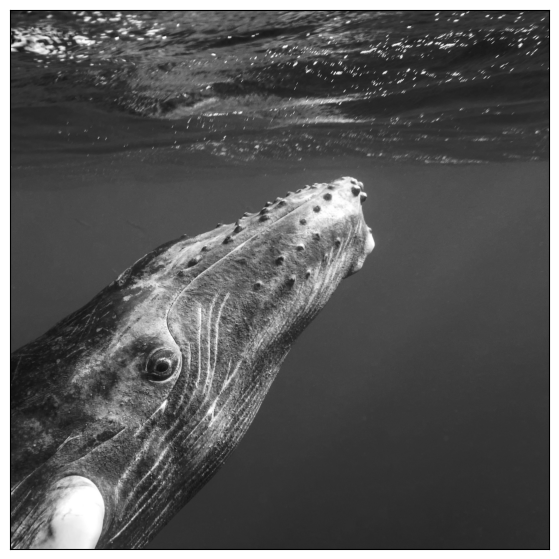}};
   		 \spy on (1.05, 0.65) in node [left] at (2.15, 0.5);
  	  \end{tikzpicture}
\end{minipage}
\begin{minipage}[t]{0.19\textwidth}
       \begin{tikzpicture}[spy using outlines={rectangle, white, magnification=2, size=0.7cm, connect spies}]
  	  \node[anchor=south west,inner sep=0]  at (0,0) {
\includegraphics[width=0.99\textwidth]{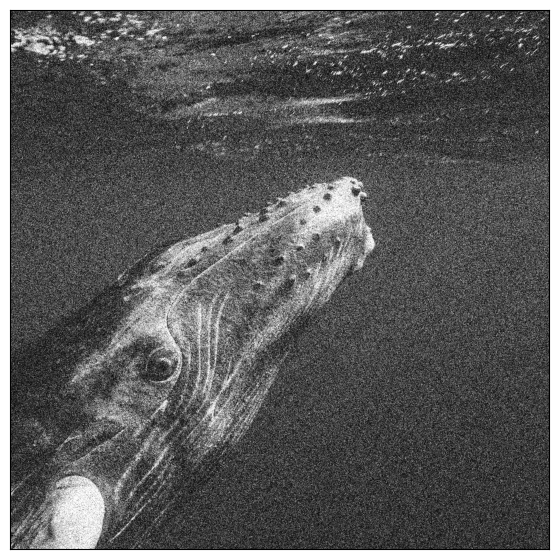}
};
   		 \spy on (1.05, 0.65) in node [left] at (2.15, 0.5);
  	  \end{tikzpicture}
\end{minipage}
\begin{minipage}[t]{0.19\textwidth}
      \begin{tikzpicture}[spy using outlines={rectangle, white, magnification=2, size=0.7cm, connect spies}]
  	  \node[anchor=south west,inner sep=0]  at (0,0) {
\includegraphics[width=0.99\textwidth]{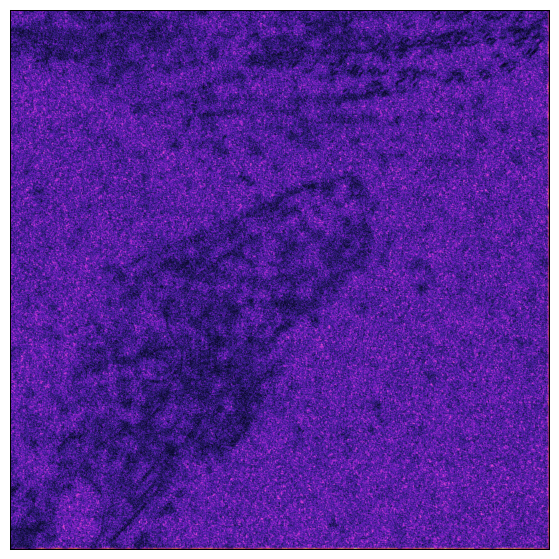}
};
   		 \spy on (1.05, 0.65) in node [left] at (2.15, 0.5);
  	  \end{tikzpicture}
\end{minipage}
\begin{minipage}[t]{0.19\textwidth}
      \begin{tikzpicture}[spy using outlines={rectangle, white, magnification=2, size=0.7cm, connect spies}]
  	  \node[anchor=south west,inner sep=0]  at (0,0) {
\includegraphics[width=0.99\textwidth]{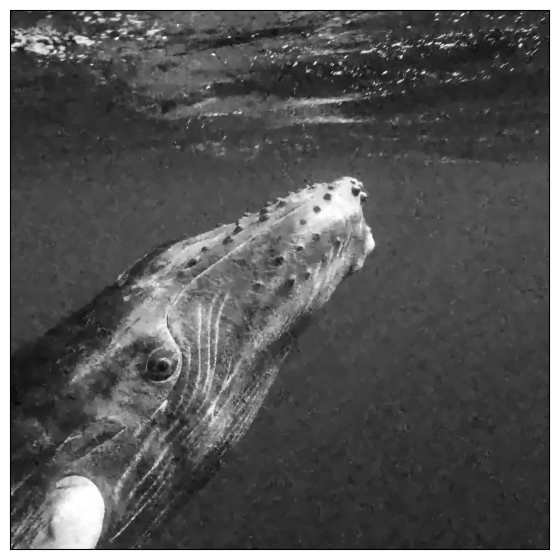}
};
   		 \spy on (1.05, 0.65) in node [left] at (2.15, 0.5);
  	  \end{tikzpicture}
\end{minipage}
\begin{minipage}[t]{0.19\textwidth}
      \begin{tikzpicture}[spy using outlines={rectangle, white, magnification=2, size=0.7cm, connect spies}]
  	  \node[anchor=south west,inner sep=0]  at (0,0) {
\includegraphics[width=0.99\textwidth]{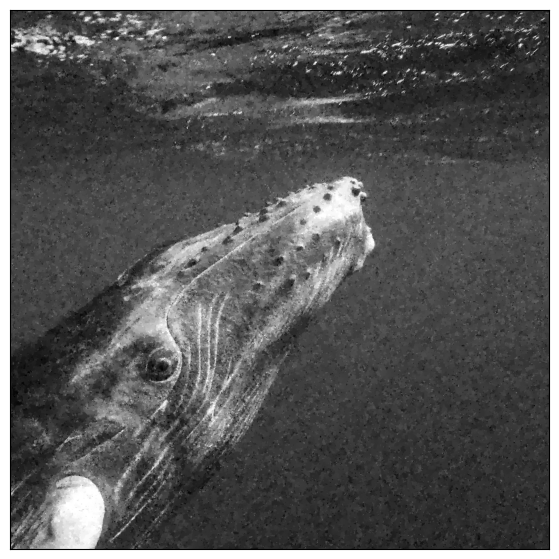}
};
   		 \spy on (1.05, 0.65) in node [left] at (2.15, 0.5);
  	  \end{tikzpicture}
\end{minipage}

\begin{minipage}[t]{0.19\textwidth}
\centering
{\footnotesize Ground truth \\ image $u_{\mathrm{true}}$}
\end{minipage}
\begin{minipage}[t]{0.19\textwidth}
\centering
{\footnotesize Noisy image \textcolor{red}{$f_{1}$} \\ (Realisation \textcolor{red}{one})\\ {\tiny PSNR=13.99, SSIM=0.12}}
\end{minipage}
\begin{minipage}[t]{0.19\textwidth}
\centering
{\footnotesize Estimated \textcolor{red}{$\Lambda_{1}$}  \\  from \textcolor{red}{$f_{1}$}}
\end{minipage}
\begin{minipage}[t]{0.19\textwidth}
\centering
{\footnotesize TV-\textcolor{red}{$\Lambda_{1}$} denoising \\ of  \textcolor{red}{$f_{1}$}\\ {\tiny PSNR=27.16, SSIM=0.77}}
\end{minipage}
\begin{minipage}[t]{0.19\textwidth}
\centering
{\footnotesize TV-\textcolor{blue}{$\Lambda_{2}$} denoising \\ of  \textcolor{red}{$f_{1}$}\\ {\tiny PSNR=24.76, SSIM=0.54}}
\end{minipage}
\vspace{0.5em}

\begin{minipage}[t]{0.19\textwidth}
\phantom{\includegraphics[width=0.99\textwidth]{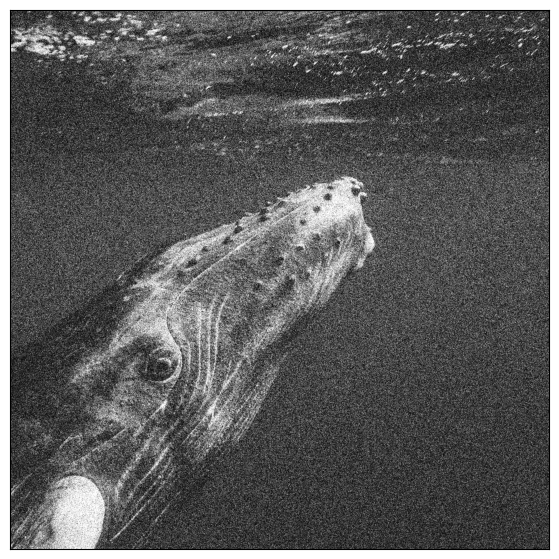}}
\end{minipage}
\begin{minipage}[t]{0.19\textwidth}
    \begin{tikzpicture}[spy using outlines={rectangle, white, magnification=2, size=0.7cm, connect spies}]
  	  \node[anchor=south west,inner sep=0]  at (0,0) {
\includegraphics[width=0.99\textwidth]{figures/results/denoising/denoising_whale_2026_08_24/noisy2.png}
};
   		 \spy on (1.05, 0.65) in node [left] at (2.15, 0.5);
  	  \end{tikzpicture}
\end{minipage}
\begin{minipage}[t]{0.19\textwidth}
   \begin{tikzpicture}[spy using outlines={rectangle, white, magnification=2, size=0.7cm, connect spies}]
  	  \node[anchor=south west,inner sep=0]  at (0,0) {
\includegraphics[width=0.99\textwidth]{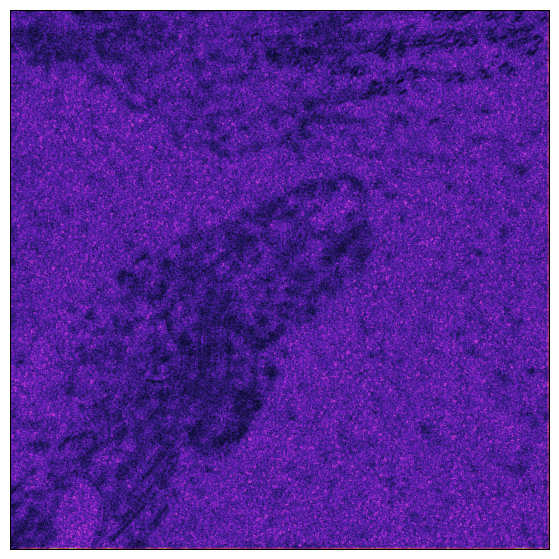}
};
   		 \spy on (1.05, 0.65) in node [left] at (2.15, 0.5);
  	  \end{tikzpicture}
\end{minipage}
\begin{minipage}[t]{0.19\textwidth}
   \begin{tikzpicture}[spy using outlines={rectangle, white, magnification=2, size=0.7cm, connect spies}]
  	  \node[anchor=south west,inner sep=0]  at (0,0) {
\includegraphics[width=0.99\textwidth]{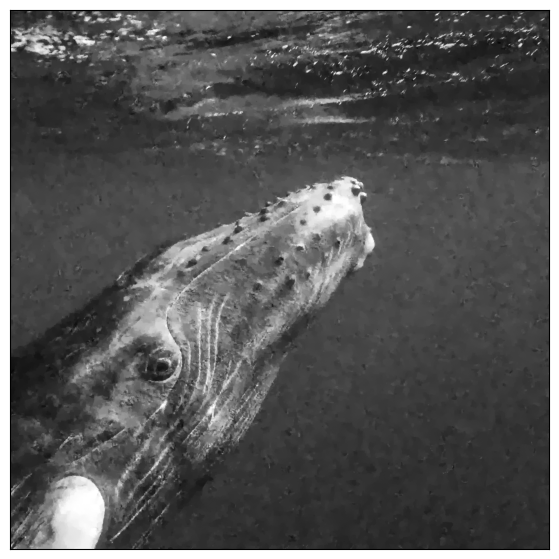}
};
   		 \spy on (1.05, 0.65) in node [left] at (2.15, 0.5);
  	  \end{tikzpicture}
\end{minipage}
\begin{minipage}[t]{0.19\textwidth}
   \begin{tikzpicture}[spy using outlines={rectangle, white, magnification=2, size=0.7cm, connect spies}]
  	  \node[anchor=south west,inner sep=0]  at (0,0) {
\includegraphics[width=0.99\textwidth]{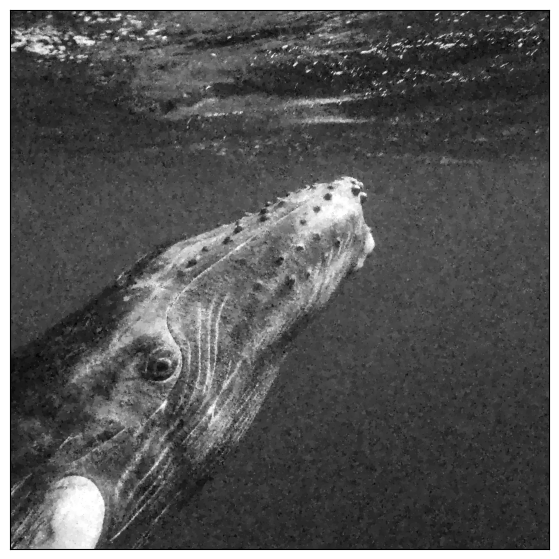}
};
   		 \spy on (1.05, 0.65) in node [left] at (2.15, 0.5);
  	  \end{tikzpicture}
\end{minipage}

\begin{minipage}[t]{0.19\textwidth}
\centering
\phantom{
{\footnotesize Clean image}}
\end{minipage}
\begin{minipage}[t]{0.19\textwidth}
\centering
{\footnotesize Noisy image \textcolor{blue}{$f_{2}$} \\ (Realisation \textcolor{blue}{two})\\{\tiny PSNR=13.97, SSIM=0.12}}
\end{minipage}
\begin{minipage}[t]{0.19\textwidth}
\centering
{\footnotesize Estimated \textcolor{blue}{$\Lambda_{2}$}  \\  from  \textcolor{blue}{$f_{2}$}}
\end{minipage}
\begin{minipage}[t]{0.19\textwidth}
\centering
{\footnotesize TV-\textcolor{blue}{$\Lambda_{2}$} denoising \\ of  \textcolor{blue}{$f_{2}$}\\ {\tiny PSNR=27.18, SSIM=0.77}}
\end{minipage}
\begin{minipage}[t]{0.19\textwidth}
\centering
{\footnotesize TV-\textcolor{red}{$\Lambda_{1}$} denoising \\ of  \textcolor{blue}{$f_{2}$}\\ {\tiny PSNR=24.75, SSIM=0.54}}
\end{minipage}

\caption{\emph{Denoising: Applying $\LL$-maps to different instances of the noise.} An experiment that shows that the parameter maps $\Lambda$ produced by the neural network-based unrolled approach of Fig.~\ref{fig:unrolled} (scheme \eqref{unrolled_network}--\eqref{supervised_learning}) are highly adapted not only to the image content but also to the specific noise instance. Here, two images with different noise instances of the same zero-mean Gaussian distribution ($\sigma^{2}=0.04$) were generated and used to infer the spatially adapted maps $\LL_{1}$ and $\LL_{2}$ for TV denoising. When these maps are cross-applied to the two noisy images, they yield considerably worse results. 
}
\label{fig:numerics_denoising}      
\end{figure}

\begin{figure}[!ht]
\centering
\begin{minipage}[t]{0.2\textwidth}
\centering
       \begin{tikzpicture}[spy using outlines={rectangle, white, magnification=2, size=0.7cm, connect spies}]
  	  \node[anchor=south west,inner sep=0]  at (0,0) {
\includegraphics[width=0.99\textwidth]{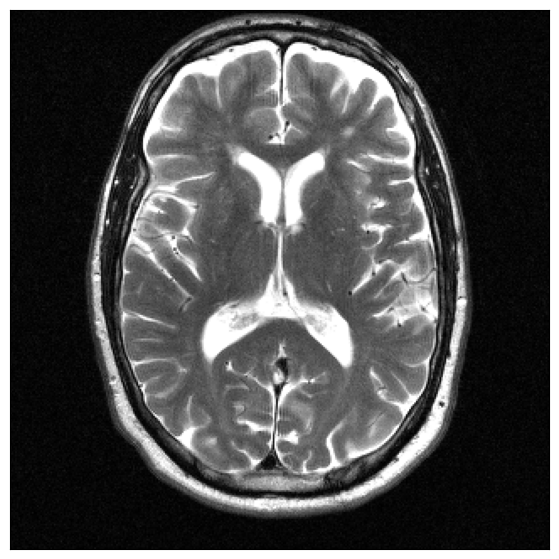}};
   		 \spy on (1.1, 0.65) in node [left] at (2.2, 0.45);
  	  \end{tikzpicture}
\end{minipage}%
\begin{minipage}[t]{0.2\textwidth}
\centering
       \begin{tikzpicture}[spy using outlines={rectangle, white, magnification=2, size=0.7cm, connect spies}]
  	  \node[anchor=south west,inner sep=0]  at (0,0) {
\includegraphics[width=0.99\textwidth]{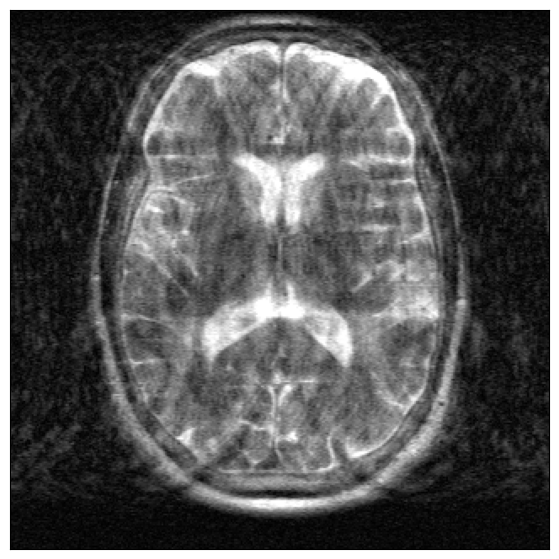}};
   		 \spy on (1.1, 0.65) in node [left] at (2.2, 0.45);
  	  \end{tikzpicture}
\end{minipage}%
\begin{minipage}[t]{0.2\textwidth}
\centering
       \begin{tikzpicture}[spy using outlines={rectangle, white, magnification=2, size=0.7cm, connect spies}]
  	  \node[anchor=south west,inner sep=0]  at (0,0) {
\includegraphics[width=0.99\textwidth]{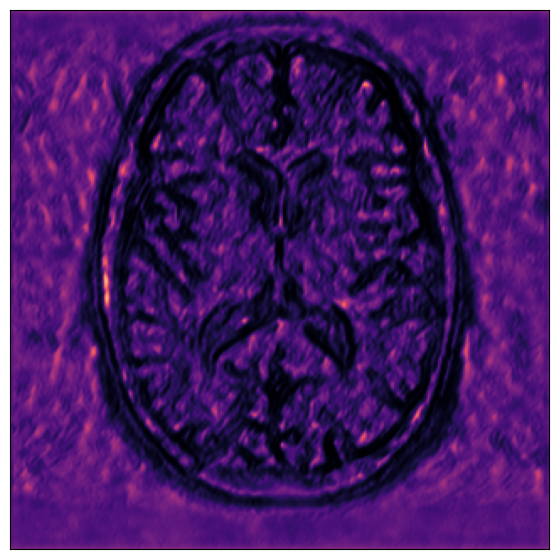}};
   		 \spy on (1.1, 0.65) in node [left] at (2.2, 0.45);
  	  \end{tikzpicture}
\end{minipage}%
\begin{minipage}[t]{0.2\textwidth}
\centering
       \begin{tikzpicture}[spy using outlines={rectangle, white, magnification=2, size=0.7cm, connect spies}]
  	  \node[anchor=south west,inner sep=0]  at (0,0) {
\includegraphics[width=0.99\textwidth]{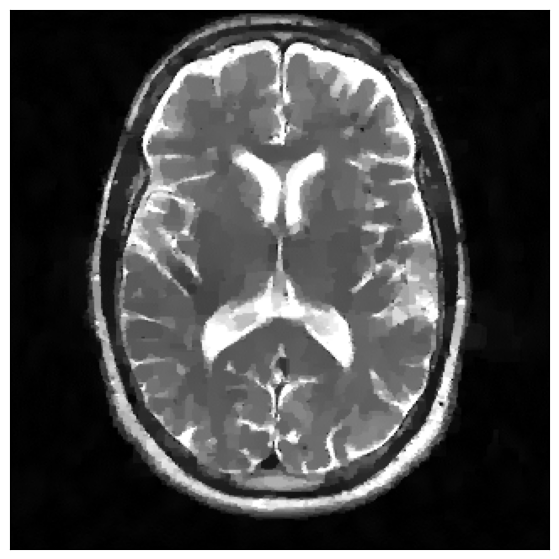}};
   		 \spy on (1.1, 0.65) in node [left] at (2.2, 0.45);
  	  \end{tikzpicture}
\end{minipage}%
\begin{minipage}[t]{0.2\textwidth}
\centering
       \begin{tikzpicture}[spy using outlines={rectangle, white, magnification=2, size=0.7cm, connect spies}]
  	  \node[anchor=south west,inner sep=0]  at (0,0) {
\includegraphics[width=0.99\textwidth]{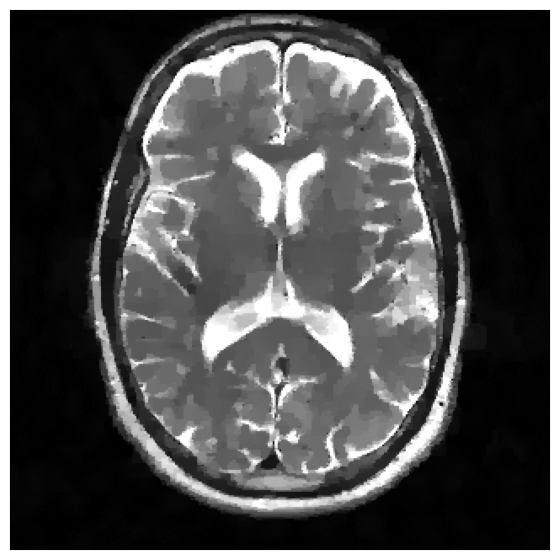}};
   		 \spy on (1.1, 0.65) in node [left] at (2.2, 0.45);
  	  \end{tikzpicture}
\end{minipage}%

\begin{minipage}[t]{0.2\textwidth}
\centering
{\footnotesize Ground truth \\ image $u_{\mathrm{true}}$}
\end{minipage}%
\begin{minipage}[t]{0.2\textwidth}
\centering
{\footnotesize Adjoint image of  \\ \textcolor{red}{$f_{1}$}$=PFu_{\mathrm{true}}+$ \textcolor{red}{$\eta_{1}$} \\ {\tiny MSE: 0.074, SSIM: 0.44}}
\end{minipage}%
\begin{minipage}[t]{0.2\textwidth}
\centering
{\footnotesize Estimated \textcolor{red}{$\Lambda_{1}$}  \\  from \textcolor{red}{$f_{1}$}}
\end{minipage}%
\begin{minipage}[t]{0.2\textwidth}
\centering
{\footnotesize TV-\textcolor{red}{$\Lambda_{1}$} reconstr.\  \\ from  \textcolor{red}{$f_{1}$}\\ {\tiny MSE=0.022, SSIM=0.72}}
\end{minipage}%
\begin{minipage}[t]{0.2\textwidth}
\centering
{\footnotesize TV-\textcolor{blue}{$\Lambda_{2}$} reconstr.\ \\ from  \textcolor{red}{$f_{1}$}\\ {\tiny MSE=0.022, SSIM=0.72}}
\end{minipage}%
\vspace{0.5em}

\begin{minipage}[t]{0.2\textwidth}
\centering
\includegraphics[width=0.99\textwidth]{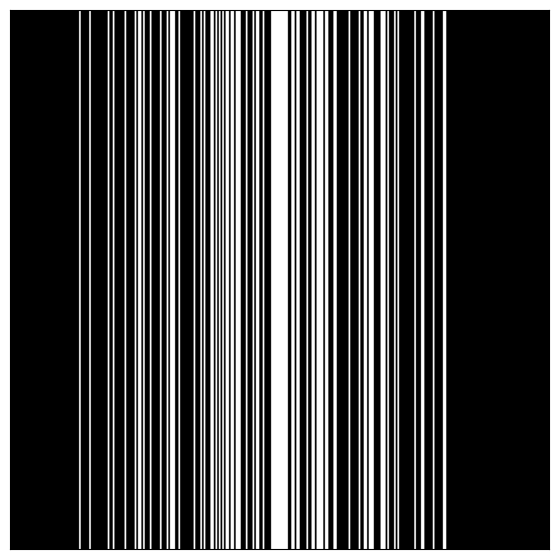}
\end{minipage}%
\begin{minipage}[t]{0.2\textwidth}
\centering
       \begin{tikzpicture}[spy using outlines={rectangle, white, magnification=2, size=0.7cm, connect spies}]
  	  \node[anchor=south west,inner sep=0]  at (0,0) {
\includegraphics[width=0.99\textwidth]{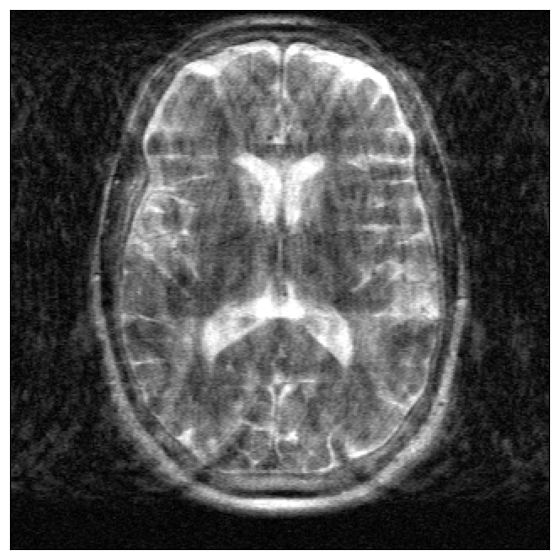}};
   		 \spy on (1.1, 0.65) in node [left] at (2.2, 0.45);
  	  \end{tikzpicture}
\end{minipage}%
\begin{minipage}[t]{0.2\textwidth}
\centering
       \begin{tikzpicture}[spy using outlines={rectangle, white, magnification=2, size=0.7cm, connect spies}]
  	  \node[anchor=south west,inner sep=0]  at (0,0) {
\includegraphics[width=0.99\textwidth]{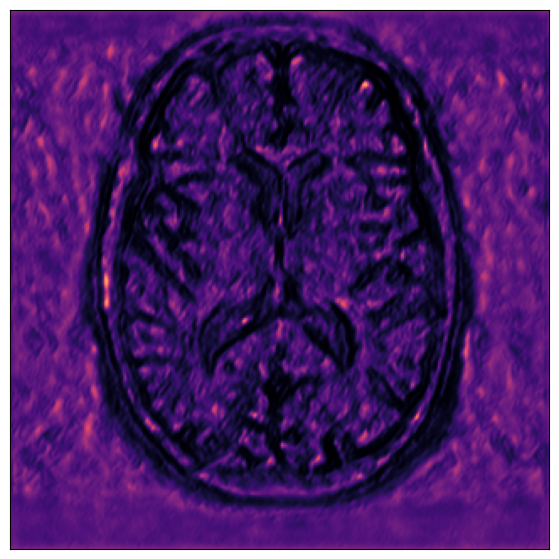}};
   		 \spy on (1.1, 0.65) in node [left] at (2.2, 0.45);
  	  \end{tikzpicture}
\end{minipage}%
\begin{minipage}[t]{0.2\textwidth}
\centering
       \begin{tikzpicture}[spy using outlines={rectangle, white, magnification=2, size=0.7cm, connect spies}]
  	  \node[anchor=south west,inner sep=0]  at (0,0) {
\includegraphics[width=0.99\textwidth]{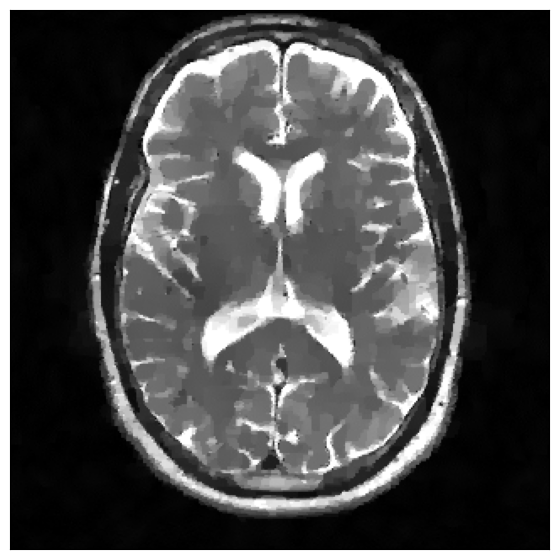}};
   		 \spy on (1.1, 0.65) in node [left] at (2.2, 0.45);
  	  \end{tikzpicture}
\end{minipage}%
\begin{minipage}[t]{0.2\textwidth}
\centering
       \begin{tikzpicture}[spy using outlines={rectangle, white, magnification=2, size=0.7cm, connect spies}]
  	  \node[anchor=south west,inner sep=0]  at (0,0) {
\includegraphics[width=0.99\textwidth]{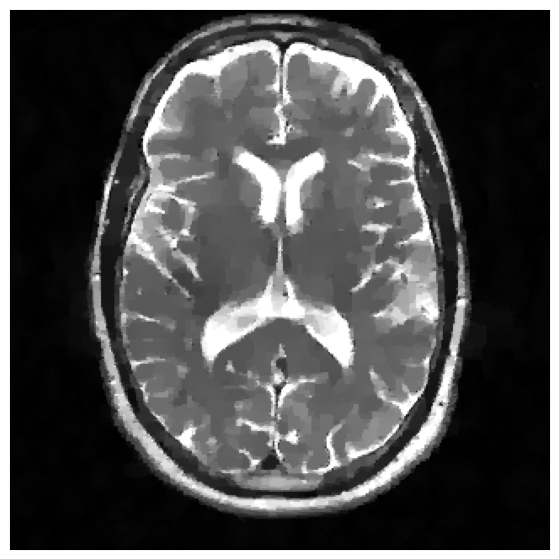}};
   		 \spy on (1.1, 0.65) in node [left] at (2.2, 0.45);
  	  \end{tikzpicture}
\end{minipage}%

\begin{minipage}[t]{0.2\textwidth}
\centering
{\footnotesize Undersampling mask $P$}
\end{minipage}%
\begin{minipage}[t]{0.2\textwidth}
\centering
{\footnotesize Adjoint image of  \\ \textcolor{blue}{$f_{2}$}$=PFu_{\mathrm{true}}+$ \textcolor{blue}{$\eta_{2}$}  \\ {\tiny MSE: 0.074, SSIM: 0.45}}
\end{minipage}%
\begin{minipage}[t]{0.2\textwidth}
\centering
{\footnotesize Estimated \textcolor{blue}{$\Lambda_{2}$}  \\  from \textcolor{blue}{$f_{2}$}}
\end{minipage}%
\begin{minipage}[t]{0.2\textwidth}
\centering
{\footnotesize TV-\textcolor{blue}{$\Lambda_{2}$} reconstr.\  \\ from  \textcolor{blue}{$f_{2}$}\\ {\tiny MSE=0.022, SSIM=0.72}}
\end{minipage}%
\begin{minipage}[t]{0.2\textwidth}
\centering
{\footnotesize TV-\textcolor{red}{$\Lambda_{1}$} reconstr.\  \\ from  \textcolor{blue}{$f_{2}$}\\ {\tiny MSE=0.022, SSIM=0.73}}
\end{minipage}%

\caption{
\emph{MRI reconstruction: Applying $\LL$-maps to data created with different instances of the noise under the same undersampling mask.}
Series of MRI reconstructions under a common sampling mask but with two different noise instances of the same Gaussian distribution added to the $k$-space data. Spatially varying regularisation maps are produced by the neural network-based unrolled approach of Fig.~\ref{fig:unrolled} (scheme \eqref{unrolled_network}--\eqref{supervised_learning}). Unlike the denoising case of Fig.~\ref{fig:numerics_denoising}, the cross application of the $\LL$-maps yields comparable reconstructions.
}
\label{fig:numerics_mri_fixed_mask_different_noise}      
\end{figure}

\begin{figure}[!ht]
\centering
\begin{minipage}[t]{0.2\textwidth}
\centering
       \begin{tikzpicture}[spy using outlines={rectangle, white, magnification=2, size=0.7cm, connect spies}]
  	  \node[anchor=south west,inner sep=0]  at (0,0) {
\includegraphics[width=0.99\textwidth]{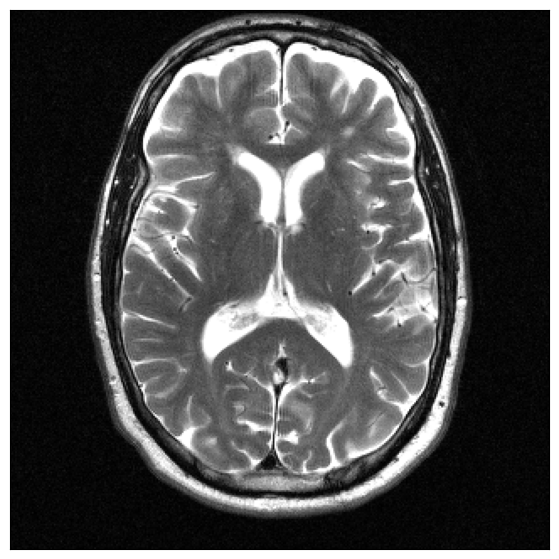}};
   		 \spy on (1.1, 0.65) in node [left] at (2.2, 0.45);
  	  \end{tikzpicture}
\end{minipage}%
\begin{minipage}[t]{0.2\textwidth}
\centering
       \begin{tikzpicture}[spy using outlines={rectangle, white, magnification=2, size=0.7cm, connect spies}]
  	  \node[anchor=south west,inner sep=0]  at (0,0) {
\includegraphics[width=0.99\textwidth]{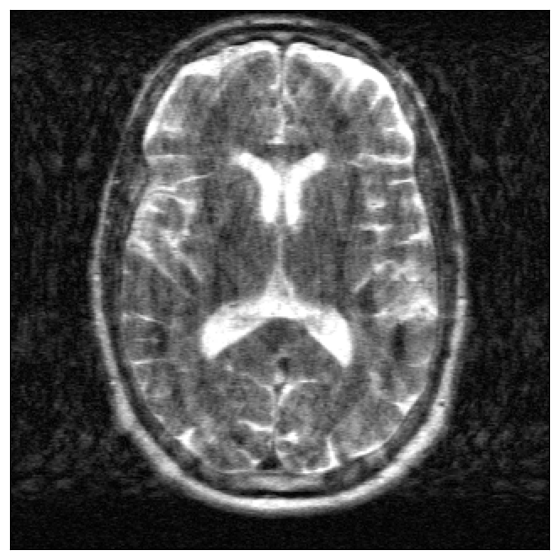}};
   		 \spy on (1.1, 0.65) in node [left] at (2.2, 0.45);
  	  \end{tikzpicture}
\end{minipage}%
\begin{minipage}[t]{0.2\textwidth}
\centering
       \begin{tikzpicture}[spy using outlines={rectangle, white, magnification=2, size=0.7cm, connect spies}]
  	  \node[anchor=south west,inner sep=0]  at (0,0) {
\includegraphics[width=0.99\textwidth]{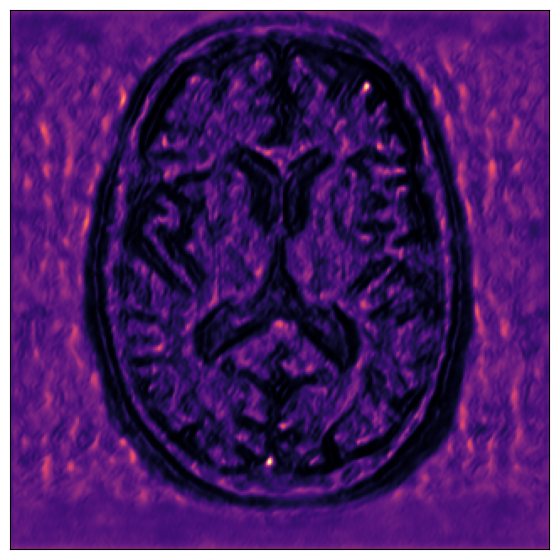}};
   		 \spy on (1.1, 0.65) in node [left] at (2.2, 0.45);
  	  \end{tikzpicture}
\end{minipage}%
\begin{minipage}[t]{0.2\textwidth}
\centering
       \begin{tikzpicture}[spy using outlines={rectangle, white, magnification=2, size=0.7cm, connect spies}]
  	  \node[anchor=south west,inner sep=0]  at (0,0) {
\includegraphics[width=0.99\textwidth]{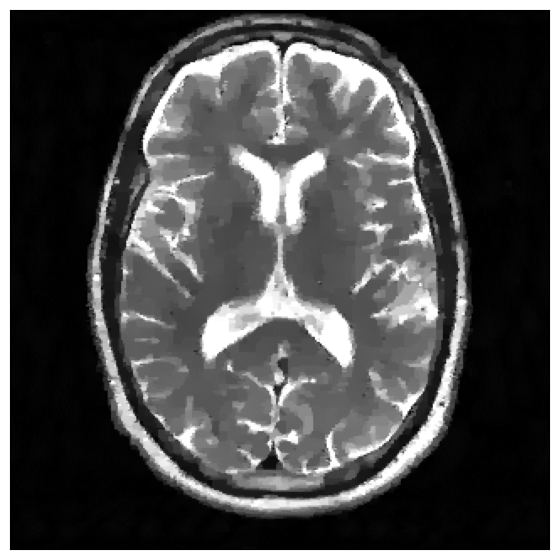}};
   		 \spy on (1.1, 0.65) in node [left] at (2.2, 0.45);
  	  \end{tikzpicture}
\end{minipage}%
\begin{minipage}[t]{0.2\textwidth}
\centering
       \begin{tikzpicture}[spy using outlines={rectangle, white, magnification=2, size=0.7cm, connect spies}]
  	  \node[anchor=south west,inner sep=0]  at (0,0) {
\includegraphics[width=0.99\textwidth]{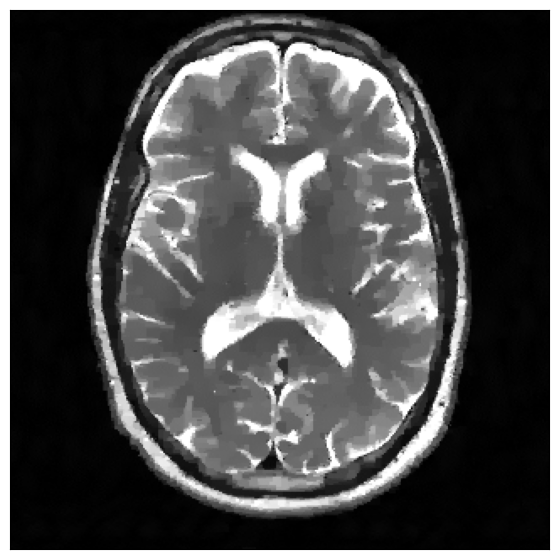}};
   		 \spy on (1.1, 0.65) in node [left] at (2.2, 0.45);
  	  \end{tikzpicture}
\end{minipage}%

\begin{minipage}[t]{0.2\textwidth}
\centering
{\footnotesize Ground truth \\ image $u_{\mathrm{true}}$}
\end{minipage}%
\begin{minipage}[t]{0.2\textwidth}
\centering
{\footnotesize Adjoint image of  \\ \textcolor{red}{$f_{1}$}$=\textcolor{red}{P_{1}}Fu_{\mathrm{true}}+$ $\eta$  \\ {\tiny MSE: 0.051, SSIM: 0.51}}
\end{minipage}%
\begin{minipage}[t]{0.2\textwidth}
\centering
{\footnotesize Estimated \textcolor{red}{$\Lambda_{1}$}  \\  from \textcolor{red}{$f_{1}$}}
\end{minipage}%
\begin{minipage}[t]{0.2\textwidth}
\centering
{\footnotesize TV-\textcolor{red}{$\Lambda_{1}$} reconstr.\  \\ from  \textcolor{red}{$f_{1}$}\\ {\tiny MSE=0.018, SSIM=0.76}}
\end{minipage}%
\begin{minipage}[t]{0.2\textwidth}
\centering
{\footnotesize TV-\textcolor{blue}{$\Lambda_{2}$} reconstr.\ \\ from  \textcolor{red}{$f_{1}$}\\ {\tiny MSE=0.017, SSIM=0.77}}
\end{minipage}%
\vspace{0.5em}

\begin{minipage}[t]{0.2\textwidth}
\centering
\includegraphics[width=0.99\textwidth]{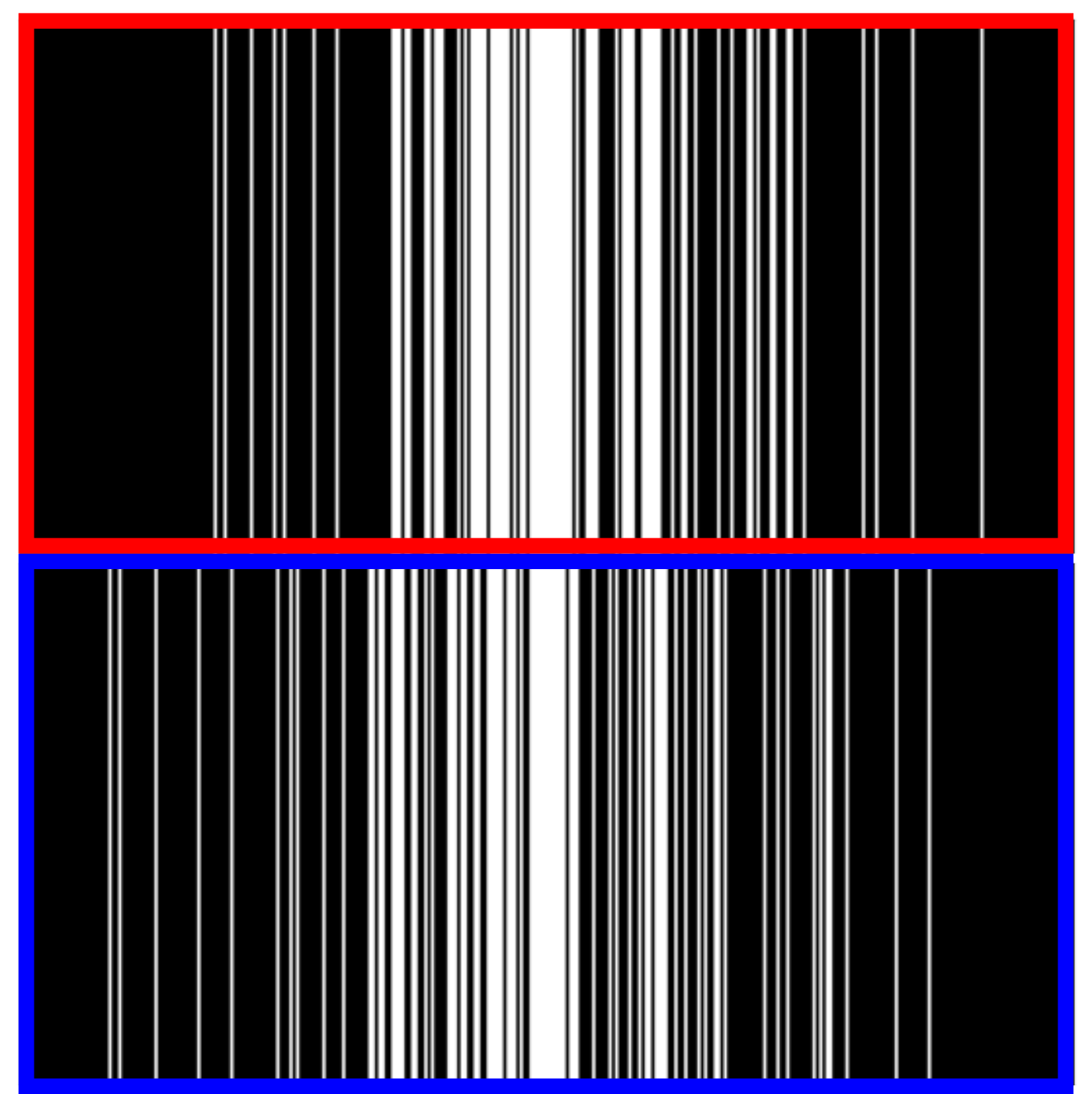}
\end{minipage}%
\begin{minipage}[t]{0.2\textwidth}
\centering
       \begin{tikzpicture}[spy using outlines={rectangle, white, magnification=2, size=0.7cm, connect spies}]
  	  \node[anchor=south west,inner sep=0]  at (0,0) {
\includegraphics[width=0.99\textwidth]{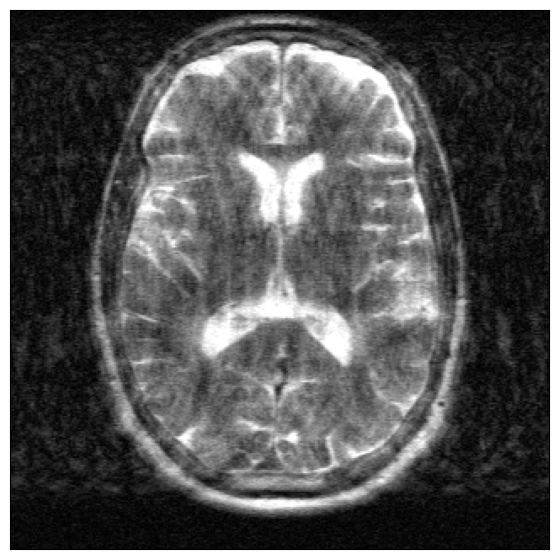}};
   		 \spy on (1.1, 0.65) in node [left] at (2.2, 0.45);
  	  \end{tikzpicture}
\end{minipage}%
\begin{minipage}[t]{0.2\textwidth}
\centering
       \begin{tikzpicture}[spy using outlines={rectangle, white, magnification=2, size=0.7cm, connect spies}]
  	  \node[anchor=south west,inner sep=0]  at (0,0) {
\includegraphics[width=0.99\textwidth]{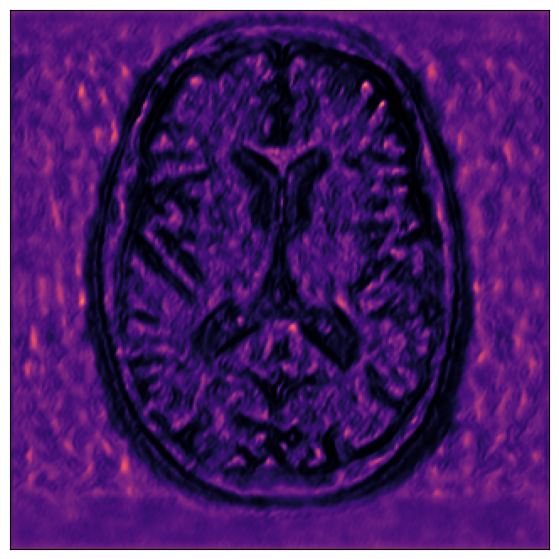}};
   		 \spy on (1.1, 0.65) in node [left] at (2.2, 0.45);
  	  \end{tikzpicture}
\end{minipage}%
\begin{minipage}[t]{0.2\textwidth}
\centering
       \begin{tikzpicture}[spy using outlines={rectangle, white, magnification=2, size=0.7cm, connect spies}]
  	  \node[anchor=south west,inner sep=0]  at (0,0) {
\includegraphics[width=0.99\textwidth]{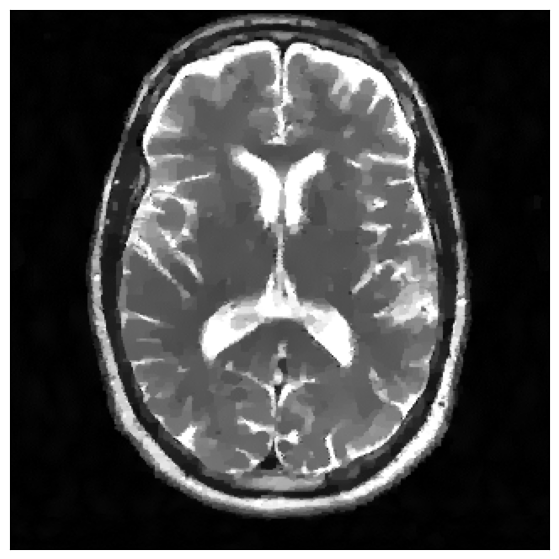}};
   		 \spy on (1.1, 0.65) in node [left] at (2.2, 0.45);
  	  \end{tikzpicture}
\end{minipage}%
\begin{minipage}[t]{0.2\textwidth}
\centering
       \begin{tikzpicture}[spy using outlines={rectangle, white, magnification=2, size=0.7cm, connect spies}]
  	  \node[anchor=south west,inner sep=0]  at (0,0) {
\includegraphics[width=0.99\textwidth]{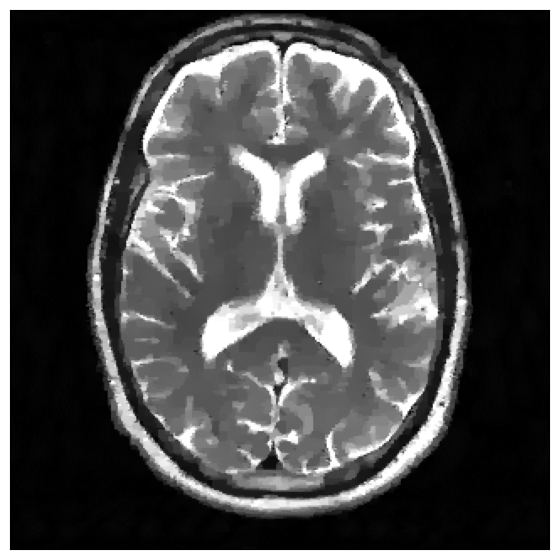}};
   		 \spy on (1.1, 0.65) in node [left] at (2.2, 0.45);
  	  \end{tikzpicture}
\end{minipage}%

\begin{minipage}[t]{0.2\textwidth}
\centering
{\footnotesize Undersampling masks \\ \textcolor{red}{$P_{1}$} and  \textcolor{blue}{$P_{2}$}}
\end{minipage}%
\begin{minipage}[t]{0.2\textwidth}
\centering
{\footnotesize Adjoint image of  \\ \textcolor{blue}{$f_{2}$}$=\textcolor{blue}{P_{2}}Fu_{\mathrm{true}}+$ $\eta$   \\ {\tiny MSE: 0.061, SSIM: 0.49}}
\end{minipage}%
\begin{minipage}[t]{0.2\textwidth}
\centering
{\footnotesize Estimated \textcolor{blue}{$\Lambda_{2}$}  \\  from \textcolor{blue}{$f_{2}$}}
\end{minipage}%
\begin{minipage}[t]{0.2\textwidth}
\centering
{\footnotesize TV-\textcolor{blue}{$\Lambda_{2}$} reconstr.\  \\ from  \textcolor{blue}{$f_{2}$}\\ {\tiny MSE=0.019, SSIM=0.74}}
\end{minipage}%
\begin{minipage}[t]{0.2\textwidth}
\centering
{\footnotesize TV-\textcolor{red}{$\Lambda_{1}$} reconstr.\  \\ from  \textcolor{blue}{$f_{2}$}\\ {\tiny MSE=0.018, SSIM=0.76}}
\end{minipage}%
\caption{
\emph{MRI reconstruction: Applying $\LL$-maps to data created with different undersampling masks under the same instance of the noise.} 
Series of MRI reconstructions under a common noise instance but with two different undersampling masks of the $k$-space data of the same rate. Spatially varying regularisation maps are produced by the neural network-based unrolled approach of Fig.~\ref{fig:unrolled} (scheme \eqref{unrolled_network}--\eqref{supervised_learning}). Unlike the denoising case of Fig.~\ref{fig:numerics_denoising}, the cross application of the $\LL$-maps yields comparable reconstructions as in the case of Fig.~\ref{fig:numerics_mri_fixed_mask_different_noise}.
}
\label{fig:numerics_mri_different_masks_fixed_noise}      
\end{figure}

\begin{figure}[!ht]
\centering
\begin{minipage}[t]{0.2\textwidth}
\centering
\begin{tikzpicture}[spy using outlines={rectangle, white, magnification=2, size=0.7cm, connect spies}]
  	  \node[anchor=south west,inner sep=0]  at (0,0) {
\includegraphics[width=0.99\textwidth]{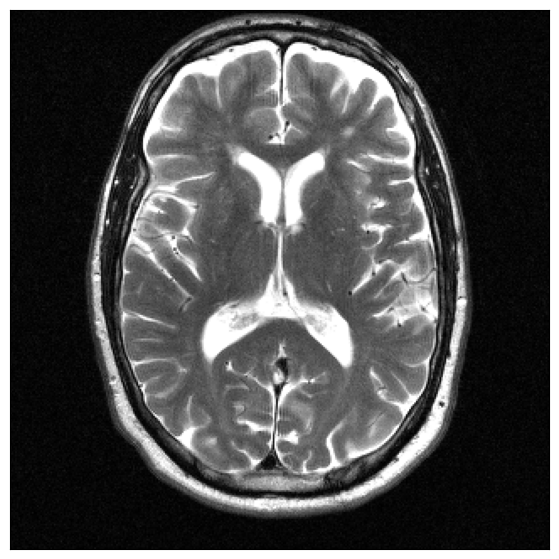}};
   		 \spy on (1.1, 0.65) in node [left] at (2.2, 0.45);
  	  \end{tikzpicture}
\end{minipage}%
\begin{minipage}[t]{0.2\textwidth}
\centering
       \begin{tikzpicture}[spy using outlines={rectangle, white, magnification=2, size=0.7cm, connect spies}]
  	  \node[anchor=south west,inner sep=0]  at (0,0) {
\includegraphics[width=0.99\textwidth]{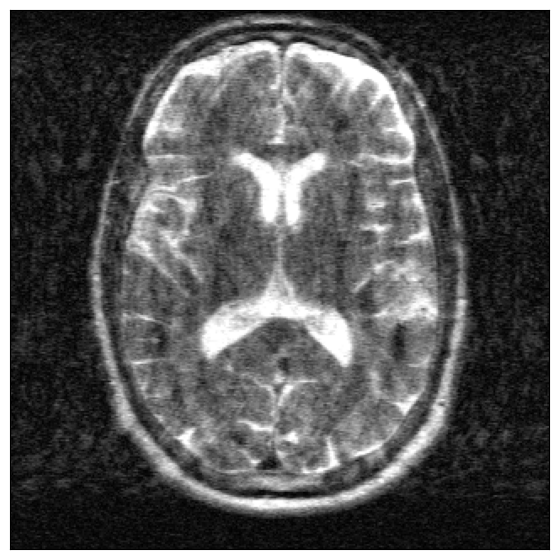}};
   		 \spy on (1.1, 0.65) in node [left] at (2.2, 0.45);
  	  \end{tikzpicture}
\end{minipage}%
\begin{minipage}[t]{0.2\textwidth}
\centering
       \begin{tikzpicture}[spy using outlines={rectangle, white, magnification=2, size=0.7cm, connect spies}]
  	  \node[anchor=south west,inner sep=0]  at (0,0) {
\includegraphics[width=0.99\textwidth]{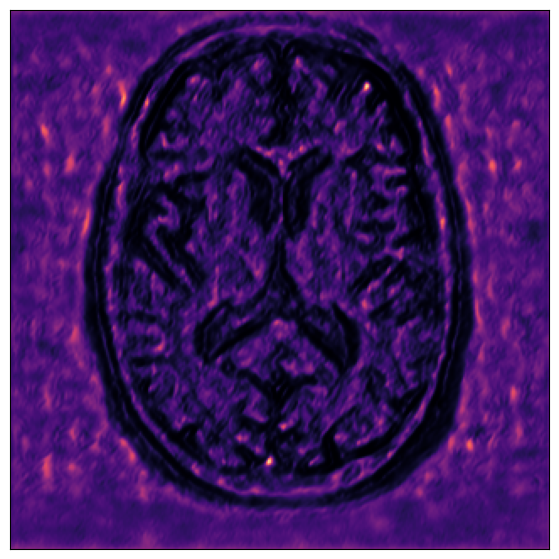}};
   		 \spy on (1.1, 0.65) in node [left] at (2.2, 0.45);
  	  \end{tikzpicture}
\end{minipage}%
\begin{minipage}[t]{0.2\textwidth}
\centering
       \begin{tikzpicture}[spy using outlines={rectangle, white, magnification=2, size=0.7cm, connect spies}]
  	  \node[anchor=south west,inner sep=0]  at (0,0) {
\includegraphics[width=0.99\textwidth]{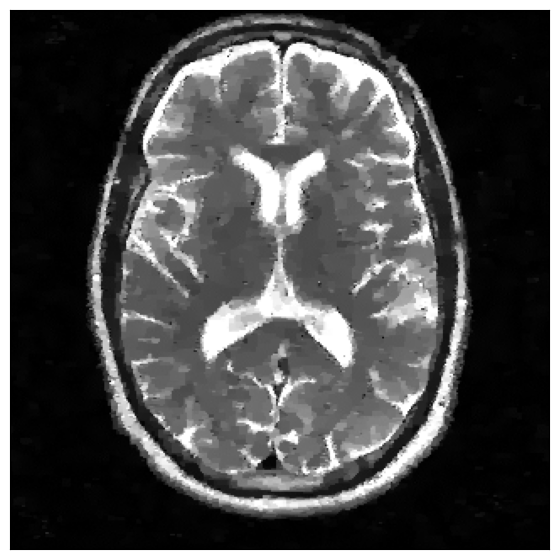}};
   		 \spy on (1.1, 0.65) in node [left] at (2.2, 0.45);
  	  \end{tikzpicture}
\end{minipage}%
\begin{minipage}[t]{0.2\textwidth}
\centering
       \begin{tikzpicture}[spy using outlines={rectangle, white, magnification=2, size=0.7cm, connect spies}]
  	  \node[anchor=south west,inner sep=0]  at (0,0) {
\includegraphics[width=0.99\textwidth]{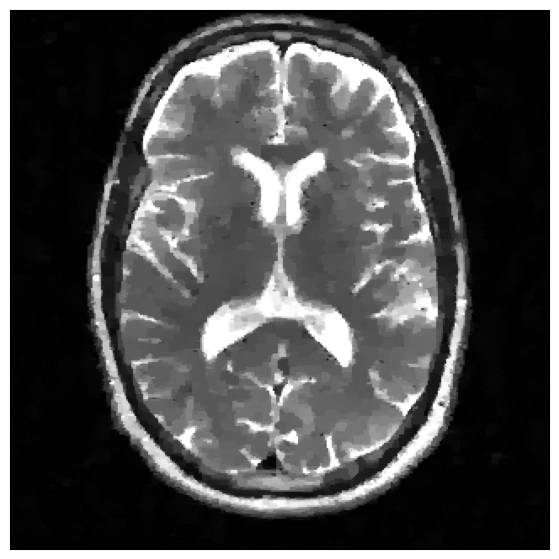}};
   		 \spy on (1.1, 0.65) in node [left] at (2.2, 0.45);
  	  \end{tikzpicture}
\end{minipage}%

\begin{minipage}[t]{0.2\textwidth}
\centering
{\footnotesize Ground truth \\ image $u_{\mathrm{true}}$}
\end{minipage}%
\begin{minipage}[t]{0.2\textwidth}
\centering
{\footnotesize Adjoint image of  \\ \textcolor{red}{$f_{1}$}$=\textcolor{red}{P_{1}}Fu_{\mathrm{true}}+$ \textcolor{red}{$\eta_{2}$}  \\ {\tiny MSE: 0.058, SSIM: 0.47}}
\end{minipage}%
\begin{minipage}[t]{0.2\textwidth}
\centering
{\footnotesize Estimated \textcolor{red}{$\Lambda_{1}$}  \\  from \textcolor{red}{$f_{1}$}}
\end{minipage}%
\begin{minipage}[t]{0.2\textwidth}
\centering
{\footnotesize TV-\textcolor{red}{$\Lambda_{1}$} reconstr.\  \\ from  \textcolor{red}{$f_{1}$}\\ {\tiny MSE=0.025, SSIM=0.76}}
\end{minipage}%
\begin{minipage}[t]{0.2\textwidth}
\centering
{\footnotesize TV-\textcolor{blue}{$\Lambda_{2}$} reconstr.\ \\ from  \textcolor{red}{$f_{1}$}\\ {\tiny MSE=0.020, SSIM=0.75}}
\end{minipage}%
\vspace{0.5em}

\begin{minipage}[t]{0.2\textwidth}
\centering
\includegraphics[width=0.99\textwidth]{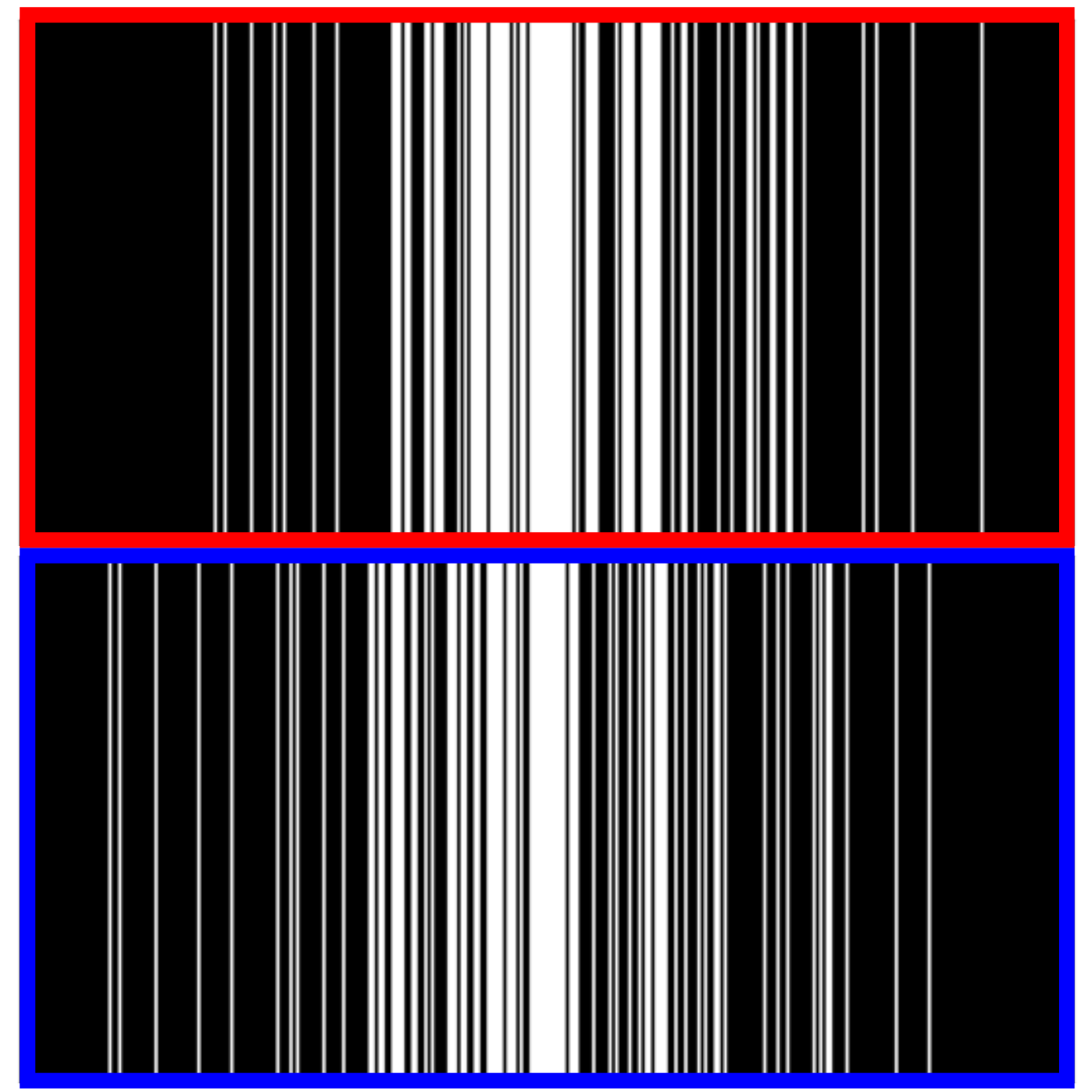}
\end{minipage}%
\begin{minipage}[t]{0.2\textwidth}
\centering
       \begin{tikzpicture}[spy using outlines={rectangle, white, magnification=2, size=0.7cm, connect spies}]
  	  \node[anchor=south west,inner sep=0]  at (0,0) {
\includegraphics[width=0.99\textwidth]{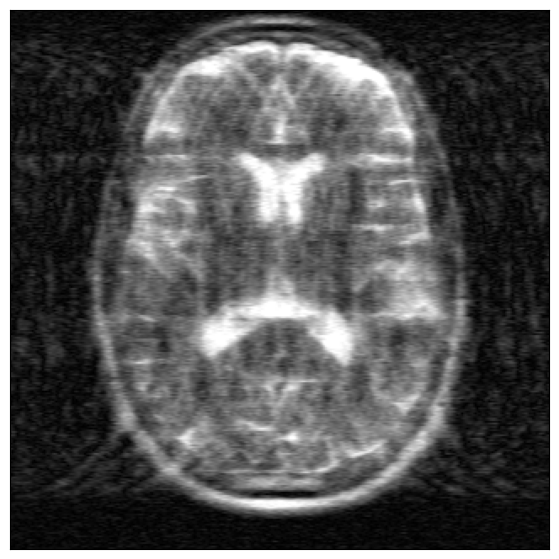}};
   		 \spy on (1.1, 0.65) in node [left] at (2.2, 0.45);
  	  \end{tikzpicture}
\end{minipage}%
\begin{minipage}[t]{0.2\textwidth}
\centering
       \begin{tikzpicture}[spy using outlines={rectangle, white, magnification=2, size=0.7cm, connect spies}]
  	  \node[anchor=south west,inner sep=0]  at (0,0) {
\includegraphics[width=0.99\textwidth]{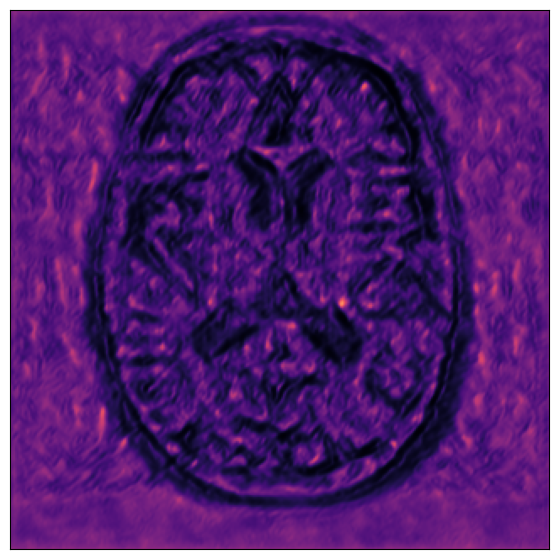}};
   		 \spy on (1.1, 0.65) in node [left] at (2.2, 0.45);
  	  \end{tikzpicture}
\end{minipage}%
\begin{minipage}[t]{0.2\textwidth}
\centering
       \begin{tikzpicture}[spy using outlines={rectangle, white, magnification=2, size=0.7cm, connect spies}]
  	  \node[anchor=south west,inner sep=0]  at (0,0) {
\includegraphics[width=0.99\textwidth]{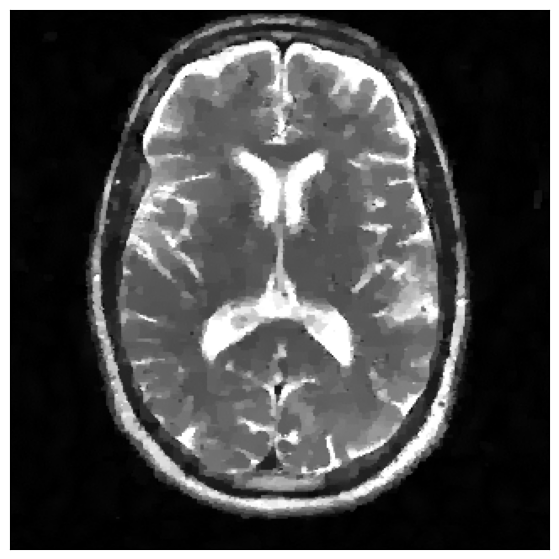}};
   		 \spy on (1.1, 0.65) in node [left] at (2.2, 0.45);
  	  \end{tikzpicture}
\end{minipage}%
\begin{minipage}[t]{0.2\textwidth}
\centering
       \begin{tikzpicture}[spy using outlines={rectangle, white, magnification=2, size=0.7cm, connect spies}]
  	  \node[anchor=south west,inner sep=0]  at (0,0) {
\includegraphics[width=0.99\textwidth]{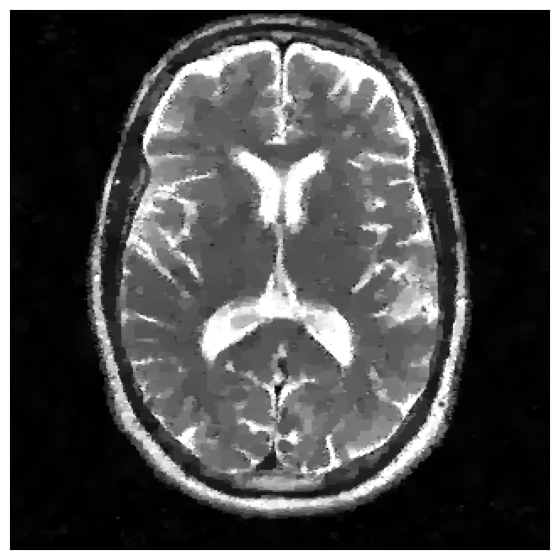}};
   		 \spy on (1.1, 0.65) in node [left] at (2.2, 0.45);
  	  \end{tikzpicture}
\end{minipage}%

\begin{minipage}[t]{0.2\textwidth}
\centering
{\footnotesize Undersampling masks \\ \textcolor{red}{$P_{1}$} and  \textcolor{blue}{$P_{2}$}}
\end{minipage}%
\begin{minipage}[t]{0.2\textwidth}
\centering
{\footnotesize Adjoint image of  \\ \textcolor{blue}{$f_{2}$}$=\textcolor{blue}{P_{2}}Fu_{\mathrm{true}}+$ \textcolor{blue}{$\eta_{2}$}   \\ {\tiny MSE: 0.075, SSIM: 0.41}}
\end{minipage}%
\begin{minipage}[t]{0.2\textwidth}
\centering
{\footnotesize Estimated \textcolor{blue}{$\Lambda_{2}$}  \\  from \textcolor{blue}{$f_{2}$}}
\end{minipage}%
\begin{minipage}[t]{0.2\textwidth}
\centering
{\footnotesize TV-\textcolor{blue}{$\Lambda_{2}$} reconstr.\  \\ from  \textcolor{blue}{$f_{2}$}\\ {\tiny MSE=0.023, SSIM=0.72}}
\end{minipage}%
\begin{minipage}[t]{0.2\textwidth}
\centering
{\footnotesize TV-\textcolor{red}{$\Lambda_{1}$} reconstr.\  \\ from  \textcolor{blue}{$f_{2}$}\\ {\tiny MSE=0.028, SSIM=0.73}}
\end{minipage}%
\caption{
\emph{MRI reconstruction: Applying $\LL$-maps to data created with different undersampling masks under two different instances of the noise.} 
Series of MRI reconstructions under two different noise instances of the same Gaussian distribution and with two different undersampling masks of the $k$-space data of the same rate. Spatially varying regularisation maps are produced by the neural network-based unrolled approach of Fig.~\ref{fig:unrolled} (scheme \eqref{unrolled_network}--\eqref{supervised_learning}). Unlike the denoising case of Fig.~\ref{fig:numerics_denoising}, the cross application of the $\LL$-maps yields comparable reconstructions as in the case of Fig.~\ref{fig:numerics_mri_fixed_mask_different_noise} and Fig.~\ref{fig:numerics_mri_different_masks_fixed_noise}.
}
\label{fig:numerics_mri_different_masks_different_noise}      
\end{figure}

In this section, we provide some numerical results using the aforementioned approach,  focusing on the regularity and the structure of regularisation parameters, also in connection with the quality of the reconstructed images. Thus, we do not get into details regarding the architecture and the training of the unrolled scheme \eqref{unrolled_network}--\eqref{supervised_learning}, for which we refer to the previously referenced papers.

Firstly, we show in Fig.~\ref{fig:denoising_regularity} what regularisation weights of high and low regularity mean in the discrete setting in practice (TV denoising). On one hand, we provide an example of a weight $\Lambda$ of high regularity produced with the bilevel optimisation approach of \cite{hintermuller2017optimal} and the corresponding denoising result. Note that this approach employs an $H^{1}$-regularisation for the weight. We compare this with the low-regularity weight and the denoised result produced by the neural network-unrolled algorithmic approach discussed in the previous section. Note that no additional regularity for the weight is imposed here, except for strict positivity by means of a softplus activation function. We observe a significantly superior reconstruction both quantitatively and qualitatively due to the higher adaptation of the low-regularity $\Lambda$.

Next, we examine up to what degree the neural network-inferred weights adapt to the image structure as well as to the noise itself. We do this investigation both for a denoising and an MRI reconstruction task. To produce these results, the TV-network implementations taken from \href{https://github.com/koflera/LearningL1NormsWeights4Synthesis}{\color{darkblue}{https://github.com/koflera/LearningL1NormsWeights4Synthesis}} were adapted for the task of image denoising and accelerated MRI, and appropriately retrained on the DIV2K  \cite{agustsson2017ntire} and fastMRI brain dataset \cite{zbontar2018fastmri}, respectively.
In Fig.~\ref{fig:numerics_denoising} we show a series of denoising results. There, a clean image $u_{\mathrm{true}}$, which is taken from the DIV2K dataset \cite{agustsson2017ntire}, is corrupted by two instances of Gaussian noise $\eta_{1}$, $\eta_{2}$ drawn from the same zero-mean Gaussian distribution of $\sigma^{2}=0.04$. This results to two noisy images $f_{1}$ and $f_{2}$ respectively. For each noisy version, we apply the unrolled scheme \eqref{unrolled_network}--\eqref{supervised_learning} with spatially varying TV as a regulariser. This produces two spatially varying regularisation parameters $\LL_{1}=\mathrm{NET}_{\theta}(f_{1})$, $\LL_{2}=\mathrm{NET}_{\theta}(f_{2})$  and two corresponding reconstructions $u_{1}:=u_{\LL_{1}}$, $u_{2}:=u_{\LL_{2}}$, see the third and fourth column of Fig.~\ref{fig:numerics_denoising} respectively. We observe that  $u_{1}$ and $u_{2}$ are very similar, both visually and quantitatively in terms of PSNR. The two corresponding regularisation maps are both highly detailed, but upon closer inspection they have remarkably different fine-scale details, which suggests that they have not only adapted to the image structure but to the specific instances of the noise. In order to further examine this, we perform a cross-application of these maps by solving a TV denoising problem for $f_{1}$ (respectively $f_{2}$) but with $\LL_{2}$ (respectively $\LL_{1}$) as regularisation parameters this time. These denoising results are shown in the last column of Fig.~\ref{fig:numerics_denoising}, where it can be clearly seen that these are significantly inferior compared to $u_{1}$ and $u_{2}$.  This confirms that each regularisation map is highly adapted to the image and the noise instance from which it was inferred, and it is not suitable to be used for denoising of the same image under a different noise instance. 

In Figs.~\ref{fig:numerics_mri_fixed_mask_different_noise}, \ref{fig:numerics_mri_different_masks_fixed_noise} and \ref{fig:numerics_mri_different_masks_different_noise}, we perform analogous experiments  for MRI reconstruction. Here the forward operator $A:= PF$ consists of an undersampling mask $P$ of the Fourier coefficients $F$ of the image at a fixed acceleration factor of $R=4$. Here, we perform three experiments, one in Fig.~\ref{fig:numerics_mri_fixed_mask_different_noise} where we keep the undersampling mask fixed and create data with two different instances of the noise, i.e.\ $f_{i}=PFu_{\mathrm{true}}+\eta_{i}$, $i=1,2$, one in Fig.~\ref{fig:numerics_mri_fixed_mask_different_noise} where the noise is fixed and we create data with two different undersampling masks of the same sampling rate,  i.e.\ $f_{i}=P_{i}Fu_{\mathrm{true}}+\eta$, $i=1,2$, and one in Fig.~\ref{fig:numerics_mri_different_masks_different_noise}, where we consider both two instances of the noise and two instances of undersampling masks,  i.e.\ $f_{i}=P_{i}Fu_{\mathrm{true}}+\eta_{i}$, $i=1,2$. In each case, we apply the scheme \eqref{unrolled_network}--\eqref{supervised_learning} for both datasets for TV denoising, and then we cross-apply the two maps as in the previous denoising case. Note that the noise realisations are always zero-mean Gaussian with $\sigma^{2}=0.0025$. The undersampling masks $P$ were generated using a Gaussian variable density sampling pattern implemented in \texttt{MRpro} \cite{zimmermann2025mrpro_arxiv, zimmermann2025mrpro}. The target images used for the retrospective generation of the Fourier measurements are examples of brain MR images taken from the fastMRI dataset \cite{zbontar2018fastmri}, which were obtained from the multi-coil Fourier data using \texttt{MRpro}.

As we see from all Figs.~\ref{fig:numerics_mri_fixed_mask_different_noise}, \ref{fig:numerics_mri_different_masks_fixed_noise} and \ref{fig:numerics_mri_different_masks_different_noise},  the inferred regularisation maps are more regular than the denoising ones, which indicates that they are less overfitted to the instances of the noise and the undersampling masks. Indeed, the cross-application of these maps to $f_{1}$ and $f_{2}$ yields very similar reconstructions, again in contrast to what we observed in denoising. This indicates that the adaptation of the weights to the specific instances of the noise and the masks does not occur here. One possible explanation is that these maps are learned from $A^{\ast}f_{i}$ instead of $f_{i}$ directly,  in which the fine details of the noise have been dampened; see the second columns of the figures above.

Finally, we mention that even though here, for simplicity, we only show  TV reconstructions, similar conclusions can be derived for the TGV ones; see, for instance, \cite{bilevelTGV, Wu_2025}. In particular, in \cite{Wu_2025} it was also observed that the TGV maps are also highly detailed and especially in denoising they seem to be highly adapted to the noise.

\section{Conclusions}

Several remarks and directions for future work can be drawn from the present discussion.\\
While the use of regularisation weights of low regularity is without doubt advantageous in practice when compared to the more regular ones, their effect to the structure of the reconstructions requires further investigation. A less regular weight function than a continuous one might be responsible for the creation of new discontinuities to a larger degree, leading to spurious artifacts in the reconstructions. However, as the numerical examples for the neural network-inferred weights show, these artifacts do not seem to occur in practice. It is unclear whether this is mostly due to the noise-overfitting behaviour resulting in a reconstruction of good quality or whether the highly oscillatory weight provides some smoothing effect. In fact, it was shown in the one-dimensional denoising case \cite{analyticalaspects_2017} that if the weight function is differentiable, the minimising solutions are constant in areas where the derivative of the weight is large enough. It would be interesting to see if this is the case for areas where the weight is not differentiable everywhere but is nevertheless characterised by abrupt changes (jumps).

Drawing connections between the highly detailed weights inferred by neural networks and the results concerning the dyadic partitions of Section \ref{sec:dyadic} is another interesting direction of future work. The results in \cite{davoli2023dyadic} suggest that highly oscillatory weights are not essential for the best approximation of a function via solutions of a weighted TV problem. At first glance, this looks contradictory to the numerical experiments of Section \ref{sec:computing}. One potential reason could be that there is a fundamental difference between the problem \eqref{min_over_partitions} considered in \cite{davoli2023dyadic} and the problem \eqref{min_over_partitions_full}, which is more closely connected to the scheme \eqref{unrolled_network}--\eqref{supervised_learning} since no intermediate fixation of the weights $\lambda_{L}$ is performed. Whether a continuous refinement of the weights in \eqref{min_over_partitions_full} is beneficial or not is closely related to whether and under which conditions this problem has a solution. If there is not, it would be interesting to consider relaxed versions of \eqref{min_over_partitions_full} over some augmented space consisting of limits of piecewise constant functions in an appropriate sense, in which such a solution exists.

Finally, it is also worthy to investigate rigorously why the adaptation of the weights to the noise occurs in denoising but not in MRI. For instance, the degree of adaptation might be linked to how ill-posed the forward operator is and up to which degree its adjoint already removes the fine-scale details of the noise.

\noindent
\subsection*{Acknowledgements} We thank the organising committee of the 19th Panhellenic Conference in Mathematical Analysis for inviting us to contribute to the conference volume with the present chapter. The work of L. Calatroni was supported by the funding received from the European Research Council (ERC) Starting project MALIN under the European Union’s Horizon Europe programme (grant 101117133). 
This work was supported by the project 22HLT02 A4IM, which has received funding from the European Partnership on Metrology, co-financed by the European Union’s Horizon Europe Research and Innovation Programme and by the Participating States.
This work represents only the view of the authors. The European Commission and the other organisations are not responsible for any use that may be made of the information it contains.

 \bibliographystyle{spmpsci}
\bibliography{refs}
\end{document}